\documentclass{aa}
\usepackage{graphicx}
\usepackage{subcaption}
\usepackage{txfonts}
\usepackage{mathrsfs} % Letttres stylisées

\newcommand{\mycommand}[1]{\texttt{#1}}
\newcommand{\heracles}{\mycommand{HERACLES++}}
\newcommand{\vund}{\mycommand{V1D}}
\newcommand{\mesa}{\mycommand{MESA}}
\newcommand{\msun}{\mycommand{$M_{\odot}$}}
\newcommand{\rsun}{\mycommand{$R_{\odot}$}}
\newcommand{\nickel}{\mycommand{$^{56}$}}

\newcommand{\iso}[2]{\ensuremath{^{#1}\rm{#2}}}
\def\nifs{\iso{56}Ni}
\def\cofs{\iso{56}Co}
\def\fefs{\iso{56}Fe}
\def\kms{km~s$^{-1}$}

\begin{document}

\title{\heracles: A multidimensional Eulerian code for exascale computing}
\titlerunning{\heracles: A multidimensional Eulerian code}
%\subtitle{}

\author{L. Roussel-Hard\inst{1,}\inst{2}, E. Audit\inst{1}, L. Dessart\inst{2},  T. Padioleau\inst{1} \and Y. Wang\inst{1}}

\institute{Université Paris-Saclay, UVSQ, CNRS, CEA, Maison de la Simulation, 91191, Gif-sur-Yvette, France\\
\email{lou.roussel-hard@cea.fr}
\and
Institut d’Astrophysique de Paris, CNRS-Sorbonne université, 98 bis boulevard Arago, 75014 Paris, France
}

\date{Received XXXX; accepted YYYY}
 
\abstract
{Numerical simulations of multidimensional astrophysical fluids present considerable challenges. However, the development of exascale computing has significantly enhanced computational capabilities, motivating the development of new codes that can take full advantage of these resources. In this article, we introduce \heracles, a new hydrodynamics code with high portability, optimized for exascale machines with different architectures and running efficiently both on CPUs and GPUs. The code is Eulerian and employs a Godunov finite-volume method to solve the hydrodynamics equations, which ensures accuracy in capturing shocks and discontinuities. It includes different Riemann solvers, equations of state, and gravity solvers. It works in Cartesian and spherical coordinates, either in 1D, 2D, or 3D, and uses passive scalars to handle gases with several species. The code accepts a user-supplied heating or cooling term to treat a variety of astrophysical contexts. In addition to the usual series of benchmarking tests, we use \heracles\ to simulate the propagation of a supernova shock in a red-supergiant star envelope, from minutes after core collapse until shock emergence. In 1D, the results from \heracles\ are in agreement with those of \vund\ for the same configuration.  In 3D, the Rayleigh-Taylor instability develops and modifies the 1D picture by introducing density and composition fluctuations as well as turbulence. The focus on a wedge, rather than the full solid angle, and the ability to run \heracles\ with a large number of GPUs allow for long-term simulations of 3D supernova ejecta with a sub-degree resolution. Future developments will extend \heracles\ to a radiation-hydrodynamics code.}


\maketitle

%-------------------------------------------------------------------

\section{Introduction}

Massive stars end their lives in the gravitational collapse of their degenerate core and the formation of a compact object, possibly followed by an energetic explosion. This phenomenon is dubbed a core-collapse supernova (SN; \citealt{Baade1934}). Most of the binding energy released by the collapsing envelope is emitted as neutrinos \citep{Arnett1989}. If revived, the SN shock born at core bounce propagates through the progenitor envelope and its highly stratified composition structure. During its expansion through the progenitor envelope, hydrodynamical instabilities grow, leading to chemical mixing on small and large scales. Spatially resolved SN remnants provide direct evidence for this chemical mixing, as well as the heterogeneous density structure of SN ejecta, with well-documented examples such as SN\,1987A or Cas A \citep{Fesen2006, Abellan2017, Milisavljevic2024}. In type II SNe, which arise from the explosion of generally red-supergiant (RSG) stars, the main instability is Rayleigh-Taylor \citep{FalkArnett1972}. Its development is influenced by the post-shock neutrino-driven convection that develops as part of the explosion \citep{Kifonidis2000, Kifonidis2003}, as well as the structured convective envelopes of these extended progenitor stars \citep{Goldberg_rsg_2022, Goldberg_sbo_2022}. The nonlinear growth of the Rayleigh-Taylor instability in SN ejecta has been extensively studied (e.g., see \citealt{Chevalier1976}) or more recently \citealt{Duffell2016}. Most of these studies were initially focused on SN\,1987A \citep{Ebisuzaki1989, Muller1991, Fryxell1991, Hachisu1994} and have been extended in recent years to type II SNe in general (e.g., \citealt{Wongwathanarat2015}).

When studying the development of instabilities in SN ejecta, the structure and chemical composition of the progenitor is critical. The Rayleigh-Taylor instability tends to develop at the H-He and the O-He interfaces \citep{Herant1991, Muller1991, Herant1993}, and the resulting structures take the form of fingers and mushrooms, altering the density and chemical profiles obtained in spherical symmetry (e.g., \citealt{Wongwathanarat2015}). The modeling of such fluids in 3D is computationally intensive and it is therefore desirable to develop efficient and robust numerical tools to tackle these scientific problems.

More and more computational capacities are available with the creation of new supercomputers and the development of new hardware. Exascale supercomputers operate with a massively parallel architecture and have a computing power exceeding \(10^{18}\) floating-point operations per second (FLOPS). The exascale era  offers the ability to solve large and complex problems, opening up new possibilities for research by combining unprecedented computing power and data analysis. Indeed, the computational resources rely less and less on CPUs and benefit more and more from GPUs. Practically, among the world's fastest supercomputers, more than 75\% of the available FLOPS are delivered by GPUs\footnote{Computed from www.top500.org}. GPUs have become a crucial driver of computational performance in high-performance computing (HPC) and different vendors (e.g., AMD, NVIDIA, Intel) are developing their own types of hardware, which can be used with their own programming languages (e.g., HIP/ROCm, CUDA, OneApi). To fully exploit these possibilities, scientists must develop software and numerical approaches that use the benefits of exascale hardware and ensure portability across different architectures.

To meet these challenges, and especially performance portability, we have developed the hydrodynamics code \heracles\ using the Kokkos library \citep{Trott2022}. The unified C++ code is portable to different architectures, offering a level of abstraction that enhances user-friendliness. The code uses the Message Passing Interface (MPI) to deal with distributed memory parallelism and Kokkos for shared memory computations. With this approach, the code can efficiently use most of the CPUs and GPUs available on the market, including NVIDIAs and AMDs.

The usage of Kokkos is slowly spreading and different codes have been recently developed using the Kokkos-C++ framework. For example,  IDEFIX \citep{Lesur2023} and  ARK \citep{Padioleau2019, Bourgeois2024} are fixed-grid magnetohydrodynamics (MHD) codes designed to study protoplanetary discs or convection, respectively.  Other codes, such as Athena++ \citep{Stone2020} and the new revision, K-ATHENA++ \citep{Grete2020}, or DYABLO \citep{Delorme2022}, implement an adaptive mesh refinement (AMR) framework for problems ranging from solar physics to cosmology. 

Some codes are more specifically dedicated to the study of core-collapse SNe (for a review, see \citealt{Muller_review_2020}). These codes were developed initially in 2D and in the last decade extended to 3D with different strategies and assets. Their aim is to explore the neutrino-driven explosion mechanism of massive stars with a variety of physical modules (e.g., neutrino transport schemes, neutrino opacities, nuclear networks, general relativity) and spatial grids (e.g. spherical polar, dendritic, Yin-Yang, Cartesian with AMR). These codes include PROMETHEUS and variants \citep{Fryxell1991, Muller1991, Rampp_2002, Wongwathanarat_2010, Wongwathanarat_2013, Wongwathanarat2015, Wongwathanarat_2017, Bollig_2021, Gabler_2021}, CHIMERA \citep{Bruenn2020, Mezzacappa_2020}, FORNAX \citep{Skinner2019, Burrows_2019, Vartanyan_2019, Burrows_2020}, COCONOT-FMT \citep{Muller_2019}, or FLASH \citep{Couch_2015, Connor_2018}. TESS and variants \citep{Duffell2011, Duffell_2013, Duffell2016, Mandal2023} use a variety of expanding grids (moving Voronoi or Cartesian meshes) to study the ejecta as they turn into SN remnants at late times. 

In our group, we have developed and used the radiation-hydrodynamics code HERACLES \citep{Gonzalez2007} for the study of SN ejecta. The focus has not been the explosion mechanism but instead the evolution of the SN ejecta on a longer timescale. We have studied in 3D the impact of ejecta structure (i.e., clumping) on type II SN light curves \citep{Dessart_2019}. We have also studied the interaction of SN ejecta with circumstellar material, both in spherical symmetry \citep{Dessart_2015} and in 2D \citep{Vlasis_16}. One asset of our approach is the post-processing of the radiation-hydrodynamics simulations with detailed nonlocal thermodynamic equilibrium (NLTE) radiative transfer for the production of not just light curves but also spectra with CMFGEN \citep{Hillier_1998, Dessart_2005, Hillier_2012, Dessart_2015, Dessart_2017}. Here, we present our current effort at switching from our previous FORTRAN-based CPU-based version of HERACLES to the new code \heracles, written in C++, making use of the Kokkos library, and adapted for exascale architectures.

In this paper, we first present the hydrodynamic equations solved for in \heracles\ as well as the physics treated (Sect. \ref{section:physics_code}). We then describe their implementation and the numerical scheme used for the resolution of the equations (Sect. \ref{section:numerical}). Some benchmarking tests are presented in Sect. \ref{section:test_hydro}, followed by a more astrophysics-oriented test in Sect. \ref{section:test_astro} with the modeling of a RSG explosion from 500\,s after the onset of the explosion until shock breakout. We present our conclusions in Sect. \ref{section:conclusion}. We also include several appendix sections to present the results from resolution tests (Appendix \ref{section:impact_resolution}), performance tests (Appendix \ref{section:performance}), as well as additional benchmarking tests (Appendix \ref{section:test_hydro_appendix}).

%--------------------------------------------------------------------
%--------------------------------------------------------------------

\section{Formulation of hydrodynamical equations and physics}
\label{section:physics_code}

In this initial version of \heracles, we solve the equations of hydrodynamics and ignore radiative transfer. We include various equations of state (EoSs), gravity, as well as the possibility of a local heating and/or cooling term. The code is designed to work in Cartesian or spherical coordinates, and in 1D, 2D, or 3D.

\subsection{Hydrodynamics}
\label{subsec:equations}

The equations of hydrodynamics in their conservative form with gravity are given by
 \begin{subequations}
    \begin{align}
         \frac{\partial \rho}{\partial t}+ \Vec{\nabla}\cdot (\rho \Vec{u})&=0 \label{eq:continuity}\\
         \frac{\partial (\rho \Vec{u})}{\partial t} + \vec{\nabla} \cdot(\rho \Vec{u}\otimes\Vec{u}+\mathbb{P})&= - \rho \Vec{g} \label{eq:momentum}\\
         \frac{\partial E}{\partial t}+ \Vec{\nabla} \cdot [ \Vec{u}(E+P)]&= -\rho \Vec{g}\cdot \Vec{u} + \Dot{e}_{\rm hc}\label{eq:energy}\\
         \frac{\partial (\rho f_\textsf{x})}{\partial t} + \Vec{\nabla}\cdot (\rho \Vec{u}f_\textsf{x})&=0 \label{eq:passive_scalar}
    ,\end{align}
    \label{eq:hydro}
\end{subequations}
where $\rho$ is the density, $\Vec{u}$ the velocity, $\mathbb{P}$ the pressure tensor, $P$ the pressure, and $\vec{g}$ the gravitational acceleration. The total energy per unit volume, $E$, is the sum of the kinetic energy, $e_{\rm k} = \rho u^2/2$, and the internal energy, $e$. The user-defined, power source term $\Dot{e}_{\rm hc}$ is used to handle a heating or a cooling process. The last equation (Eq. \ref{eq:passive_scalar}) can be added to track the distribution of elements in the course of the simulation. $f_\textsf{x}$ is the mass fraction of species ${\textsf{x}}$ and defined such that $\sum_{\textsf{x}} f_{\textsf{x}}= 1$.

\subsection{Gravity terms}
\label{subsec:gravity}

Different types of gravity are implemented in \heracles. The first option is uniform gravity, which is defined by
\begin{equation}
    \Vec{g} = \Vec{g_0}
    \label{eq:uniform_gravity}
,\end{equation}
where $\Vec{g_0}$ is the constant and uniform gravitational acceleration. It can be used in all geometries.

A second option treats the gravity from a central mass, $M_{\rm c}$, in spherical geometry. The gravitational acceleration is then given by
 \begin{equation}
    \Vec{g} = \Vec{g}(r) = -\frac{GM_{\rm c}}{r^2} \,\Vec{e_r}
    \label{eq:point_gravity}
,\end{equation}
where $G$ is the gravitational constant and $r$ is the radius.

Self-gravity of the outer layers can also be accounted for in the approximation of a spherically symmetric matter distribution. In this case, we have:
 \begin{equation}
    \Vec{g} = \Vec{g}(r, t) = -\frac{GM(r, t)}{r^2} \,\Vec{e_r}
    \label{eq:internal_gravity}
,\end{equation}
where $M(r, t)$ is the total mass within the sphere of radius $r$ at time $t$.

\subsection{Equation of state}
\label{subsec:eos}

\heracles\ can work with a generic EoS. In the present work, two EoSs are used. The first one treats the ideal gas, for which the pressure of the gas and the sound speed are given by
\begin{equation}
    P = P_g = (\gamma - 1)e \quad \textrm{ and} \quad c_s = \sqrt{\frac{\gamma \, P_g}{\rho}}
    \label{eq:eos_gas}
,\end{equation}
where $\gamma$, the adiabatic index, can be chosen by the user.

The second EoS treats together an ideal gas in equilibrium with radiation. In this case, the total internal energy and pressure are the sum of the gas and the radiative components:
 \begin{subequations}
    \begin{align}
        e = \frac{\rho k_b T}{\mu (\gamma-1)} + a_r T^4
        \label{eq:eos_gp_rad_ein}\\
         P = P_g + P_r \quad \textrm{with} \quad P_r = \frac{1}{3}\, a_rT^4
    \label{eq:eos_gp_rad}
    ,\end{align}
\end{subequations}
where $k_b$ is the Boltzman constant, $\mu$ the mean molecular weight, and $a_r$ is the radiative constant. The sound speed is then given by 
\begin{equation}
     c_s = \sqrt{\frac{\gamma / (\gamma - 1) + 20 \alpha + 16 \alpha^2}{1/(\gamma-1) + 12 \alpha} \frac{P_g}{\rho}}
    \label{eq:cs_gp_rad}
,\end{equation}
where $\alpha = P_r / P_g$. 

For this EoS, the pressure cannot be computed directly from the internal energy, so we first need to determine the temperature. This is done using the internal energy equation \ref{eq:eos_gp_rad_ein} and a Newton-Raphson approach, which converges in approximately ten iterations.
This EoS is used for simulations of stellar explosions, which are characterized by radiation-dominated conditions. This also enables the continuation with radiation hydrodynamics of an originally hydrodynamics-only simulation (the switch should be done when somewhere on the grid the conditions are ripe for photon decoupling from the gas).

\subsection{Energy source terms: Heating and cooling}
\label{subsec:heat_ni}

A term, $\Dot{e}_{\rm hc}$, may be added to the energy equation (Eq.~\ref{eq:energy}) in order to study specific astrophysical conditions.  It may be positive (heating) or negative (cooling). It allows one for example to treat radioactive decay or radiative losses. The user can define a custom function fit to the physical setup being studied.

When studying SN ejecta, unstable isotopes decay and release neutrinos, $\gamma$ rays, and positrons in the process. At present, we treat the \nifs\ decay chain, with daughter isotopes \cofs\ and \fefs\ \citep{Nadyozhin1993}. The associated term in the energy equation is given by

 \begin{equation}
     \Dot{e}_{\rm hc}(t) = \lambda\Big\{ \Big[ Q_{\rm Ni}\Big(\frac{\tau_{\rm Co}}{\tau_{\rm Ni}} -1 \Big)- Q_{\rm Co}\Big] e^{-t/\tau_{\rm Ni}}+ Q_{\rm Co}e^{-t/\tau_{\rm Co}}\Big\} \, \rho f_{{\rm Ni}}  
     \label{eq:heating_ni}
 ,\end{equation}
where $Q_{\rm Ni}=1.75$\,MeV and $Q_{\rm Co}=3.73$\,MeV are the total energy ($\gamma$-ray and positrons) emitted per decay of \nifs\ and \cofs, and $\lambda$ is given by 
  \begin{equation}
     \lambda = \frac{1}{56 m_u\,(\tau_{\rm Co} - \tau_{\rm Ni})} \, .
 \end{equation}
The lifetimes of \nifs\ and \cofs\ are given by $\tau_{\rm Ni} =$\,8.8\,d  and $\tau_{\rm Co}=$\,111.3\,d. At the present time, we assume that this decay power is fully absorbed locally.

For radioactive decay heating, this emissivity term depends only the total yield of \nifs\ and its spatial distribution (i.e., the total number of \nifs\ nuclei at all radii). However, $\Dot{e}_{\rm hc}$ may also depend on the temperature, as in the cases of nuclear burning or radiative cooling, and in this case we must iterate to find the proper $\Dot{e}_{\rm hc}$ and temperature for the corresponding EoS.

\subsection{Monitoring of internal energy in highly supersonic regions}
\label{subsec:pressure_fixe}

When the internal energy becomes too small compared to the total energy, it becomes difficult to track it precisely using the total energy conservation (Eq.\,\ref{eq:energy}). This can be an important issue when, for example, the temperature is needed. Instead, we introduce the following equation to correct its evolution:
 \begin{equation}
     \frac{\partial e}{\partial t}+ \vec{\nabla}\cdot(\Vec{u}e) + P\vec{\nabla}\cdot\Vec{u}=0   \quad
     \label{eq:pressure_fix}
 .\end{equation}
The evolution of the  internal energy is then given by 
\begin{equation}
    e(t+dt) = e(t) + \alpha \, \delta^1 e + (1-\alpha)  \, \delta^2 e   \, ,
\end{equation}
where $\delta^1 e$ is the evolution of the internal energy given by the integration of Eq.~\ref{eq:energy} and $\delta^2 e$ is determined by the integration of Eq.~\ref{eq:pressure_fix}, using the velocities from the standard hydrodynamical step.

The condition for $\alpha$ is
\begin{equation}
\alpha =\left\{
    \begin{aligned}
   & 1 \quad &\textrm{if} \quad e \ge \varepsilon \, e_k\\
   & \Big(\frac{e}{\varepsilon \, e_k}\Big)^2 \quad &\textrm{if} \quad \varepsilon \, e_k > e  > \frac{\varepsilon \, e_k}{10} \\
   & 0 \quad &\textrm{if} \quad e \le \frac{\varepsilon \, e_k}{10}
     \end{aligned}
    \right.
,\end{equation}
where $\varepsilon$ is a criterion fixed by the user (we typically choose $\varepsilon= 10^{-6}$).

%--------------------------------------------------------------------
%--------------------------------------------------------------------

\section{Numerical methods and implementation}

\label{section:numerical}

\heracles\ is designed to offer flexibility in selecting the physics modules to be used, to accommodate various numerical schemes as well as switching easily between different geometries and dimensions. We present below the implementation of a Godunov-type finite-volume method.

\subsection{Numerical discretization}
\label{subsec:discretization}

Our notation for space-time discretization is such that the position at the center of the cell is denoted by $r_i$ and the mesh nodes are given by $r_{i\pm1/2}$. The cell size is $\delta r$. The time interval between $t^n$ and $t^{n+1}$ is $\delta t$. The notation $U^n_i$ represents the average quantity at time $t^n$ for the cell centered at $r_i$. This notation is trivially extended to all geometries and in several dimensions. When using spherical coordinates, $i$ refers to the radius coordinate, $j$ to the polar-angle coordinate, and $k$ to the azimuthal-angle coordinate.

\subsubsection{Mesh}

In \heracles, the computational domain is defined as a logically Cartesian mesh that can be either in Cartesian or spherical coordinates and in 1D, 2D, or 3D. The surface  elements ($dS$) and volumes ($dV$) of cells are computed based on the chosen geometry. The grid spacing in each direction can be either regular or logarithmic.  Users also have the option to implement a custom grid spacing. It is also possible to initialize the simulation from an HDF5 file containing both the physical quantities and the grid spacing.

In addition, the code includes a moving grid feature that allows one to shift the grid along a specified coordinate axis during the simulation. This functionality allows one to track regions of interest defined by the user.
In the context of SN ejecta, the grid can be allowed to move in the radial direction in order to follow the expanding ejecta. With this option, the grid size can be reduced or resolution can be enhanced in specific regions. Furthermore, the elimination of cells in the inner ejecta reduces the constraints on the CFL conditions.

\subsubsection{Momentum equation discretization in spherical coordinates}

 Equations~\ref{eq:hydro} are written in a conservative form. However, in spherical coordinates, the tensor operator (Eq. 1b) presents geometrical source terms. We use a classical discretization method that conserves angular momentum. The azimuthal component, $\phi$, of the momentum equation is then written as:
\begin{equation}
\begin{split}
    \frac{\partial \rho u_{\phi}}{\delta t} + \frac{1}{r^2}\frac{\partial r^2 \rho u_{\phi}u_r}{\partial r}
     + \frac{1}{r\sin\theta}\frac{\partial \sin\theta \,\rho u_{\phi}u_{\theta}}{\partial\theta} 
     \\+\frac{1}{r\sin\theta} \frac{\partial (\rho u_{\phi}^2 + P_{\phi \phi})}{\partial \phi} = - \, \Big(\frac{\rho u_{\phi}u_r}{r} + \frac{\cot \theta \, \rho u_{\phi} u_{\theta}}{r}\Big)
\end{split}
.\end{equation}
The source terms are discretized as follow:
\begin{equation}
\begin{aligned}
\begin{split}
    \frac{\rho u_{\phi}u_r}{r} = \frac{r_{i+1/2} - r_{i-1/2}}{(r_{i+1/2} + r_{i-1/2}) \, dV} \Big( [\rho u_{\phi}u_r]_{i+1/2} \,dS_{i+1/2}^r \\
    +  [\rho u_{\phi}u_r]_{i-1/2}\, dS_{i-1/2}^r\Big) \\
   \frac{\cot\theta \, \rho u_{\phi} u_{\theta}}{r} = \frac{\sin\theta_{j+1/2} - \sin\theta_{j-1/2}}{(\sin\theta_{j+1/2} + \sin\theta_{j-1/2}) \, dV} \\
   \times  \Big([\rho u_{\phi} u_{\theta}]_{j+1/2} \,dS_{j+1/2}^{\theta} + [\rho u_{\phi} u_{\theta}]_{j-1/2}\, dS_{j-1/2}^{\theta} \Big)
    \label{eq:source_terms_1}
\end{split}
\end{aligned}
,\end{equation}
where $dS^{r}$ is the surface elements in the radial direction, $dS^{\theta}$ in the polar direction, and $dV$ the volume element. The values at the cell interfaces are computed using outputs from the Riemann solver used for the cell update. 

The corresponding equation for the polar component, $\theta$, is
\begin{equation}
\begin{split}
    \frac{\partial \rho u_{\theta}}{\delta t} + \frac{1}{r^2}\frac{\partial r^2 \rho u_{\theta}u_r}{\partial r} + \frac{1}{r\sin\theta}\frac{\partial \sin\theta  (\rho u_{\theta}^2 + P_{\theta\theta})}{\partial\theta}\\ +\frac{1}{r\sin\theta} \frac{\partial \rho u_{\theta}u_{\phi}}{\partial \phi}
     = - \, \Big(\frac{ \rho u_{\theta}u_r}{r} -\cot\theta\, \frac{\rho u_{\phi}^2}{r}\Big)
\end{split}
,\end{equation}
with the first source term discretized as Eq.~\ref{eq:source_terms_1} and the second:
\begin{equation}
\begin{split}
    \cot\theta\, \frac{\rho u_{\phi}^2}{r} = \frac{([\rho u_{\phi}^2]_{k+1/2} + [\rho u_{\phi}^2]_{k-1/2})\cos[(\theta_{j+1/2}+\theta_{j-1/2})/2]}{2\sin[(\theta_{j+1/2}+\theta_{j-1/2})/2]} 
    \\ \times \frac{(dS_{j+1/2}^{\theta} - dS_{j-1/2}^{\theta})}{2\, dV}
\end{split}
.\end{equation}

Other source terms are discretized in a similar manner. 

\subsection{Algorithmic steps}
\label{subsec:algorithmic_steps}

\heracles\ was designed to easily incorporate a variety of numerical schemes. In this paper, we concentrate on a Godunov-type approach \citep{Godunov1959} and more specifically on the Monotone Upstream-centered Schemes for Conservation Laws (MUSCL)-Hancock  method \citep{van_Leer1977a, van_Leer1977b, van_Leer1979, van_Leer1984, Toro2009} to solve the hydrodynamical equations. This method is a simple but practical choice for achieving second-order accuracy in numerical simulations. Its ability to accurately represent gradients and discontinuities, its robustness, and its conservation properties make it a reliable tool to treat complex flows involving strong shocks.\\

The first step of the MUSCL scheme is to reconstruct time and space extrapolated variables at the cell interfaces. The space reconstruction is done with a classical slope-limiter approach: 
\begin{equation}
    \Tilde{U}^n_L = \Tilde{U}^n_i - \frac{\delta r_i}{2}\sigma \quad \textrm{and} \quad \Tilde{U}^n_R = \Tilde{U}^n_i + \frac{\delta r_i}{2}\sigma
,\end{equation}
where $\Tilde{U}_L$ and $\Tilde{U}_R$ are the reconstructed primitive variables ($\rho, \Vec{u}, P$) at the left and right sides of the cell,  and $\sigma$ the slope defined using a slope limiter. Three different slope limiters are implemented in \heracles: Minmod \citep{Sandham1989}, Van Leer, and Van Albada \citep{Berger2005, Toro2009}. The user can choose the slope limiter at runtime and it is also possible to remain at first order. From these primitive reconstructed variables, we can obtain the reconstructed conservative variables, $U_{L,R}$. 

The time reconstructed values at the half time step are obtained in the following way:
\begin{equation}
     U^{n+1/2}_{L,R} = U^n_{L,R} + \frac{\delta t}{ 2 \delta r_i} \Big[F(U^n_{L}) - F(U^n_{R})\Big] 
,\end{equation}
where $F$ is the flux. The fluxes are computed from these space-time extrapolated values using the analytical flux function. In 3D, the reconstruction is done in a similar manner (e.g., see Sect.~16.5 of \citealt{Toro2009}).

The evolution to the next time step is made using the Godunov scheme with the Riemann solver:
\begin{equation}
    U_i^{n+1}=U_i^n+\frac{\delta t}{\delta r}\Big(\mathcal{F}_{i-1/2}^{n+1/2}-\mathcal{F}_{i+1/2}^{n+1/2}\Big)
    \label{eq:godunov}
,\end{equation}
where $\mathcal{F}_{i-1/2}^{n+1/2}$ is the flux computed with the chosen Riemann solver using the space-time extrapolated values on each side of the interface.
\heracles\ currently implements three choices of Riemann solvers: the classical HLL and HLLC solvers \citep{Toro2009} and an all-mach regime solver that is much more accurate for low Mach regions \citep{Bourgeois2024} . The user can choose the solver at runtime and the structure of the code makes it easy to implement a new one.

Three boundary conditions are implemented: periodic, null gradient, and reflective.  A specific module also allows for the implementation of user-defined boundary conditions.

\subsection{Implementation}
\label{subsec:implementation}

\heracles\, is designed with an emphasis on both performance and portability. This enables the use of a single source code that can be compiled across various architectures (GPUs, CPUs), while ensuring optimal performance on each. The code is modular, making it easy to implement and switch between different slope limiters and Riemann solvers for the flux calculation.

\subsubsection{Parallelism}

The code can leverage MPI for distributed memory parallelism and the Kokkos library \citep{Trott2022} for shared memory parallelism. The global computational domain is partitioned into equal subdomains, with each subdomain assigned to a separate MPI process. Within each subdomain, the Kokkos library is used to take advantage of shared memory parallelism.

\heracles\ has a modular structure, with each module corresponding to a specific step in the hydrodynamic scheme. A module consists of a Kokkos kernel, which is a C++ functor, and is executed using the Kokkos parallel patterns, such as \texttt{Kokkos::parallel\_for} or \texttt{Kokkos::parallel\_reduce}. For example, Eq.~\ref{eq:godunov} is implemented as:\\
\\
\texttt{Kokkos::parallel\_for(} \\
\texttt{"Parallel\_for\_example",}\\
\texttt{Kokkos::MDRangePolicy<Kokkos::Rank<3>\hspace{0.01cm}>} \\
\texttt{(\{0, 0, 0\}, \{nx, ny, nz\}),} \\
\texttt{KOKKOS\_LAMBDA(int i, int j, int k)\{} \\
\texttt{\hspace*{0.5cm} F\_L = compute\_FL(i,j,k);} \\
\texttt{\hspace*{0.5cm} F\_R = compute\_FR(i,j,k);} \\
\texttt{\hspace*{0.5cm} dv = compute\_dV(i,j,k);} \\
\texttt{\hspace*{0.5cm} U\_new(i,j,k) = U(i,j,k) + dt/dv * (F\_L-F\_R));}\\
\\
where \verb|U_new|, \verb|U|, and \verb|S| are all of the \verb|Kokkos::View| type, which is the fundamental data type used in the Kokkos library. The \verb|MDRangePolicy| specifies how to iterate over a multidimensional container.

\subsubsection{Inputs and outputs}

The input-output (I/O) operations in \heracles\ are handled by PDI,\footnote{PDI stands for Portable Data Interface. Information can be found at https://pdi.dev/master/} a library designed to decouple high-performance simulation codes from I/O management. The simulation generates a significant amount of data, which is output by PDI into HDF5 files. We can initialize a simulation by directly assigning values to the physical fields or by reading an HDF5 file containing the values for these fields. This approach allows us to restart a simulation from a specific checkpoint by loading the corresponding data file.

Another advantage of using PDI is that the I/O configuration is fully decoupled from the simulation itself. This allows us to adjust the I/O settings in a separate YAML file without the need to recompile \heracles.

%--------------------------------------------------------------------
%--------------------------------------------------------------------

\section{Benchmarking tests}
\label{section:test_hydro}

We have conducted a large number of tests to validate and benchmark \heracles. These include the standard suite of hydrodynamical tests for shocks, advection, implosion, or explosion. In  Appendix~\ref{section:test_hydro_appendix}, we present results for a subset of such tests, in particular those that were performed in 2D and 3D. In this section, we present results for the tests that have a more direct relevance to the context of multidimensional simulations of stellar explosions. Hence, we present results for a 2D Kelvin-Helmholtz instability test in Sect.~\ref{section:kh}, for a 3D Rayleigh-Taylor instability test in Sect.~\ref{section:RT_3D}, and for a 3D Sedov blast wave test in Sect.~\ref{section:sedov_3D}. We then move on to discuss a more astrophysics-oriented test in Sect.~\ref{section:test_astro} with the simulations in 1D and 3D of the propagation of a SN shock in a RSG star envelope.

\begin{figure*}[h]
   \centering
   \includegraphics[width=\textwidth]{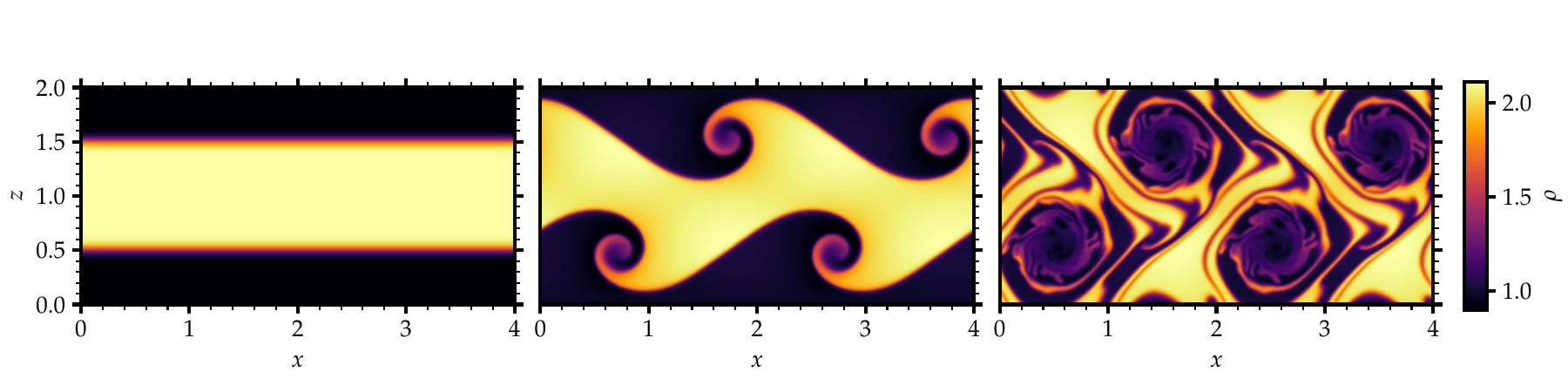}
   \caption{Color map of the density for the 2D Kelvin-Helmholtz instability test and shown at times $0$ (left), $3$ (middle) and $6$ (right). The grid resolution corresponds to $[n_x,n_z]= [2048,1024]$. (See Sect. \ref{section:kh} for discussion.)}
   \label{fig:KH}
\end{figure*}

\subsection{Two-dimensional Kelvin-Helmholtz instability test}
\label{section:kh}

The Kelvin-Helmholtz instability test is useful to investigate the development of turbulence arising from a velocity shear within a fluid. Following \cite{Lecoanet2016}, the initial conditions in the $(x,z)$ plane for the density, $\rho$, the $x$-velocity, $u_x$, and the $z$-velocity, $u_z$, are given by
\begin{equation}
     \begin{aligned}
         & \rho=1+\frac{1}{2}\Big[\tanh\Big(\frac{z-z_1}{a}\Big)-\tanh\Big(\frac{z-z_2}{a}\Big)\Big] \, , \\
         &u_x = \tanh\Big(\frac{z-z_1}{a}\Big)-\tanh\Big(\frac{z-z_2}{a}\Big) \, , \\
         &u_z = 0.01\sin(\pi x)\Big[ \exp\Big(-\frac{(z-z_1)^2}{\sigma^2} \Big) +\exp\Big(-\frac{(z-z_2)^2}{\sigma^2} \Big) \Big] \, ,
     \end{aligned}
 \end{equation}
with $a=0.05$, $\sigma=0.2$, $z_1=0.5$, and $z_2=1.5$. The initial pressure, $P_0$, is 10 and we assume an ideal gas with $\gamma=5/3$. The boundary conditions are periodic in every direction.  

\begin{figure}
   \centering
   \includegraphics[width=\columnwidth]{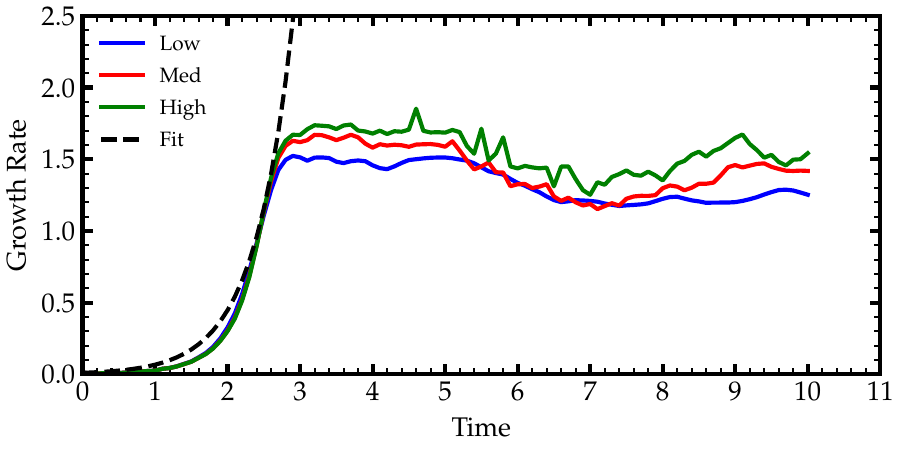}
   \caption{Instability growth rate for the 2D Kelvin-Helmholtz instability test shown in Fig.~\ref{fig:KH}. This growth rate is estimated from the evolution of the quantity  $\partial \ln u_z^{max} /\partial t$ and for three different grid choices, corresponding to ``Low'' (blue), ``Med'' (red), and ``High'' (green) resolution (i.e., $[n_x,n_z]= [512, 256]$, $[1024, 512]$, and $[2048, 1024]$). The dashed line corresponds to the best-fitting exponential to the growth rate in the linear regime, which breaks down at $t \sim 2.6$.}
   \label{fig:growth_kh}
\end{figure}

\begin{figure*}
    \centering
    \includegraphics[width=7.4cm]{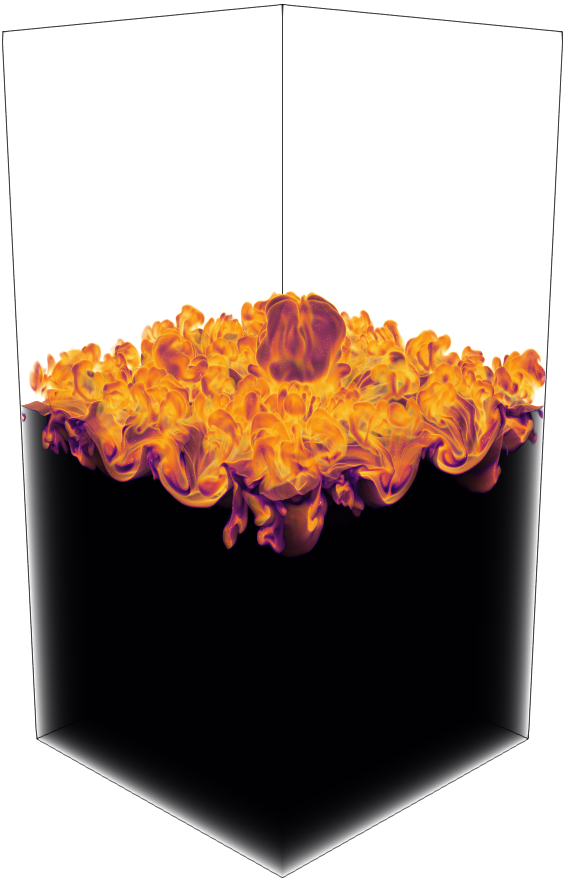} 
    \hfill
    \includegraphics[width=\columnwidth]{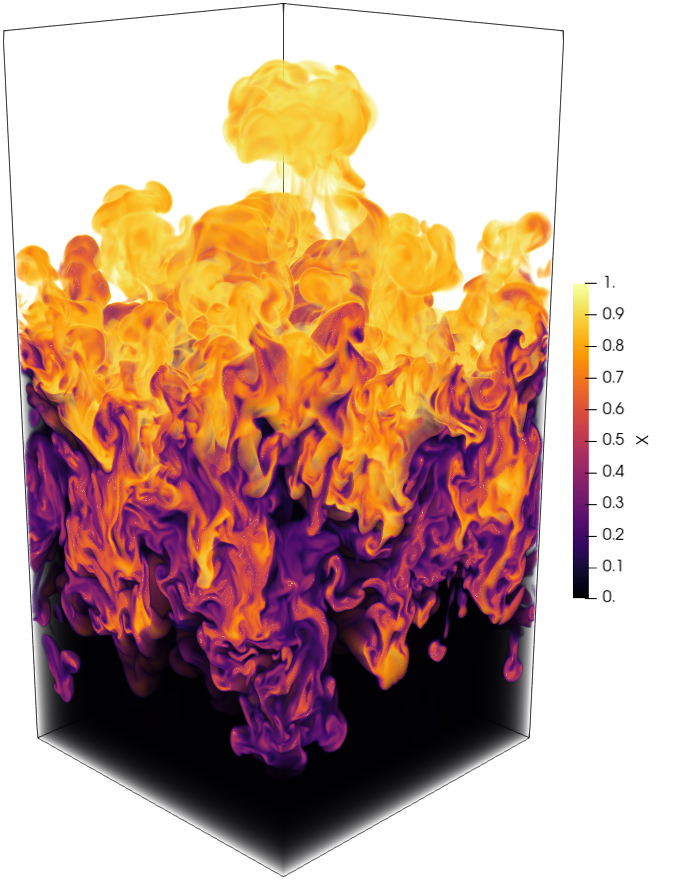}
    \caption{Passive scalar concentration, $f_\textsf{x}$, for the 3D Rayleigh-Taylor instability test at a time of $5$ (left) and $10$ (right). The Cartesian grid is characterized by $[n_x, n_y, n_z] = [512,512,1024]$. (See Sect. \ref{section:RT_3D} for discussion.)}
    \label{fig:RT_3D}
\end{figure*}

Figure~\ref{fig:KH} shows the evolution of the density at three times ($t=0, 3, 6$). The instability results in the creation of small, alternate vortices, which cascade into turbulence. In Fig.~\ref{fig:growth_kh}, we quantify the growth rate of the instability-generated structures by evaluating the evolution of the quantity $\partial \ln u_z^{max} /\partial t$, with $u_z^{max}$ the maximum velocity at each time, and for different resolutions: $512\times256$ (blue), $1024\times512$ (red), and $2048\times1024$ (green). The linear phase of the instability ends at $t\sim 2.6$ and the dashed black line represents the best exponential fitting for this phase, corresponding to $0.01 e^{1.9 t}$. The growth rate flattens after that time and remains roughly constant until the end of the simulation, largely independent of the resolution.

We can compare our results to those published by \cite{Skinner2019}. The linear evolution that they obtain ends around a time of 2.5 with a value of 1.5, which are similar to our results. Their growth rate is between 1.5 and 2, whereas ours is a little lower between 1.2 and 1.7. When comparing Fig. \ref{fig:KH} with the counterpart in their work, we observe fewer structures and a more diffusive scheme, which can explain the lower values in our growth rate. In both simulations, the resolution modestly affects the development of turbulence. Indeed, we only obtain a slight increase in the growth rate when employing a higher resolution.

\subsection{Three-dimensional Rayleigh-Taylor instability test}
\label{section:RT_3D}

The first three-dimensional test that we present is the development of the nonlinear R-T instability. The initial conditions are taken from \cite{Skinner2019}. The gravitational acceleration is constant with $g_z = -2$ and the initial density of the fluid is stratified such that:
\begin{equation}
    \rho(z)=\rho_0 \,\Big(1 - \frac{\gamma - 1}{\gamma}\frac{\rho_0 g z}{P_0}\Big)^{\frac{1}{\gamma -1}} ,\quad \rho_0= \left\{\begin{aligned}
    & \rho_h = 3 \quad \textrm{if} \quad z \ge 0 \\
    & \rho_l = 1\quad \textrm{otherwise}
        \end{aligned}
\right.
,\end{equation}
where $\gamma = 5/3$. The initial pressure is chosen to ensure that the initial conditions are in hydrostatic equilibrium:
\begin{equation}
    P= P_0\Big(\frac{\rho}{\rho_0}\Big)^\gamma \quad \textrm{with} \quad P_0 = 2\pi(\rho_h +\rho_l)gL \, .
\end{equation}

\begin{figure}
   \centering
   \includegraphics[width=\columnwidth]{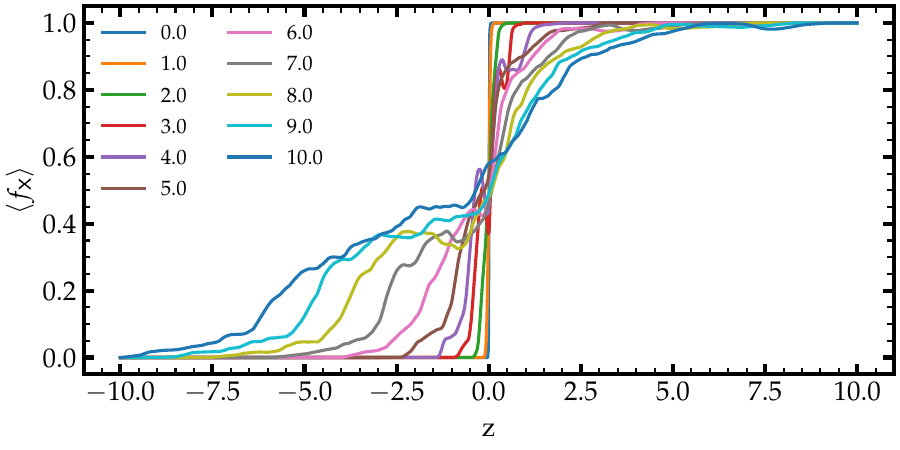}
   \caption{Multiepoch vertical profile of the horizontally averaged passive scalar $\langle f_\textsf{x}\rangle$ obtained in the 3D Rayleigh-Taylor instability test (see also Fig.~\ref{fig:RT_3D}, and Sect.\ref{section:RT_3D} for discussion).}
   \label{fig:mixing_RT_3D}
\end{figure}

At $t=0$, the interface between the two fluids, which is originally at $z=0$, is displaced by an amount, h(x,y), given by
\begin{equation}
    \begin{split}
        h(x,y)&= \frac{1}{H} \sum_{8\le k_x^2+k_y^2\le 16} \Big[ a \cos(k_x X)\cos(k_y Y)+
        b \cos(k_x X)\sin(k_y Y) 
      \\ &+ c \sin(k_x X)\cos(k_y Y) + d \sin(k_x X)\sin(k_y Y)\Big]
        \label{eq:perturb_RT}
    \end{split}
,\end{equation}
where $X=2\pi x/L$, $Y=2\pi y/L$ and $k_x, k_y$ are the wave numbers with values in the range zero to four. The coefficients, $a$, $b$, $c$, and $d$, are sampled from a uniform distribution randomly sampled from -1 to 1, excluding the boundaries. The normalized coefficient is $H=\sqrt{(1/4)(a^2 + b^2 + c^2 + d^2 )} / (h_{\rm rms})$ and $h_{\rm rms}=3 \times 10^{-4} L$.

We tracked the evolution of the fluid with a passive scalar field initialized to unity for  positive $z$ and zero otherwise. We assumed periodic boundary conditions in the $x$ and $y$ directions  and reflective conditions in the $z$ direction. The size of the domain is $x, y \in [-L/2, L/2]$ and $z \in [-L, L]$, with $L=10$. The number of grid zones in all three directions is $[n_x, n_y, n_z] = [512,512,1024]$.

Figure~\ref{fig:RT_3D} shows the evolution of the passive scalar at $t=5$ (left) and $t=10$ (right). The initial perturbation of the interface between the two fluids generates small buoyant bubbles that grow, occupying a greater fraction of the volume as time passes.

We can quantify the mixing between the two fluids by evaluating the horizontally averaged concentration of the passive scalar versus $z$ defined as:
\begin{equation}
    \langle f_\textsf{x}\rangle = \frac{1}{L^2} \int \int f_\textsf{x} \, dx dy   \, .
\end{equation}

The vertical profile of the quantity, $\langle f_\textsf{x}\rangle$, is shown at multiple epochs in Fig.~\ref{fig:mixing_RT_3D}. Initially, the interface between the two fluids at $z\sim0$ is sharp. However, as time passes, the mixing layer thickens and eventually encompasses the full height of the simulated domain. This mixing occurs on large scale and corresponds to advection of material with distinct $f_\textsf{x}$, whose average value covers from 0 to 1 over the full height. 

\subsection{Three-dimensional off-centered Sedov blast wave test}
\label{section:sedov_3D}

The test of the Sedov blast wave \citep{Sedov1946} is useful to describe the energy-conserving phase of a SN remnant. We performed this test using a grid in spherical geometry in order to test the precision of our numerical scheme for a non-Cartesian grid.  The problem consists of a stationary background medium with a uniform density, $\rho_0=1$, and a low pressure, $P_0=0.01$. For our test, the domain was a spherical wedge with $r\in[0.1, 1.3]$ and $(\theta, \phi) \in [\pi /4, 3\pi/4]$. The blast was created with a total energy peak of $E_1 = 1$ at the position $\Vec{r_0} = (r_0,\theta_0,\phi_0)=(0.7,\pi/2,\pi/2)$. The boundary conditions are transmissive in every direction, although the simulation is stopped before the blast wave reaches any boundary. The spherical polar grid counts 256 zones in all three directions.

The distance between the shock and $\Vec{r_0}$ is given analytically by
\begin{equation}
    d_{\rm sh}(t)=\xi_0 \Big(\frac{E_1\, t^2}{\rho_0}\Big)^{1/5}
    \label{eq:analytical_sedov}
,\end{equation}
with $\xi_0\approx 1.15$ for an adiabatic EoS with $\gamma=5/3$ \citep{Rosswog2022}. It therefore predicts that the initial energy injection develops into an expanding spherical blast that migrates away uniformly from $\Vec{r_0}$.

\begin{figure}
   \centering
   \includegraphics[width=7.5cm]{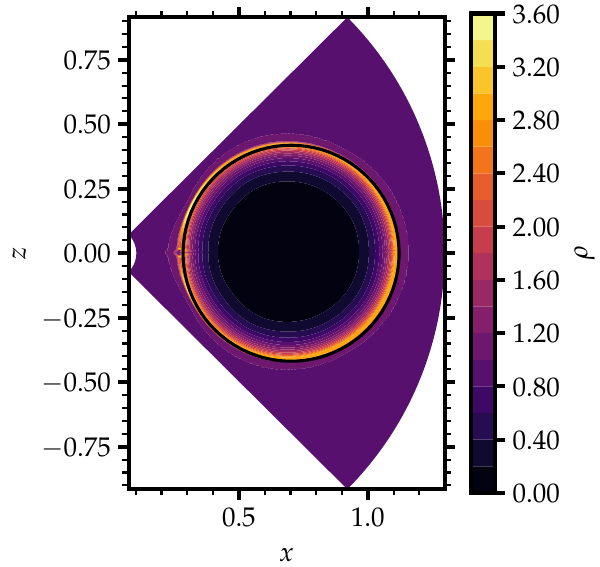}
   \caption{Meridional slice of the density for the 3D off-centered Sedov blast wave test. The time is 0.08 and the cut is done along $\phi=\pi/2$. The analytical prediction given by Eq.~\ref{eq:analytical_sedov} is shown as a black line. (See Sect. \ref{section:sedov_3D} for discussion.)}
   \label{fig:sedov_3D}
\end{figure}

Figure~\ref{fig:sedov_3D} shows a meridional slice of the density at $t= 0.08$ together with the analytical solution given by Eq.~\ref{eq:analytical_sedov}. Other slices through the 3D domain yield a similar pattern. Hence, the simulation is in good agreement with the analytical prediction, apart from a small offset in the direction where the shock is aligned with the grid. In such locations, the numerical diffusion is lower, causing a notorious problem for numerical schemes \citep{Elling2009}.

%--------------------------------------------------------------------

\section{Astrophysical test: The case of a red-supergiant star explosion and the development of Rayleigh-Taylor instability structures}

\label{section:test_astro}

%----------------------------------------------------------------------
% Preambel: context, references article pour l'evolution des progenitreur et explosions etc.
%----------------------------------------------------------------------

\begin{figure*}
    \centering
    \includegraphics[width=\textwidth]{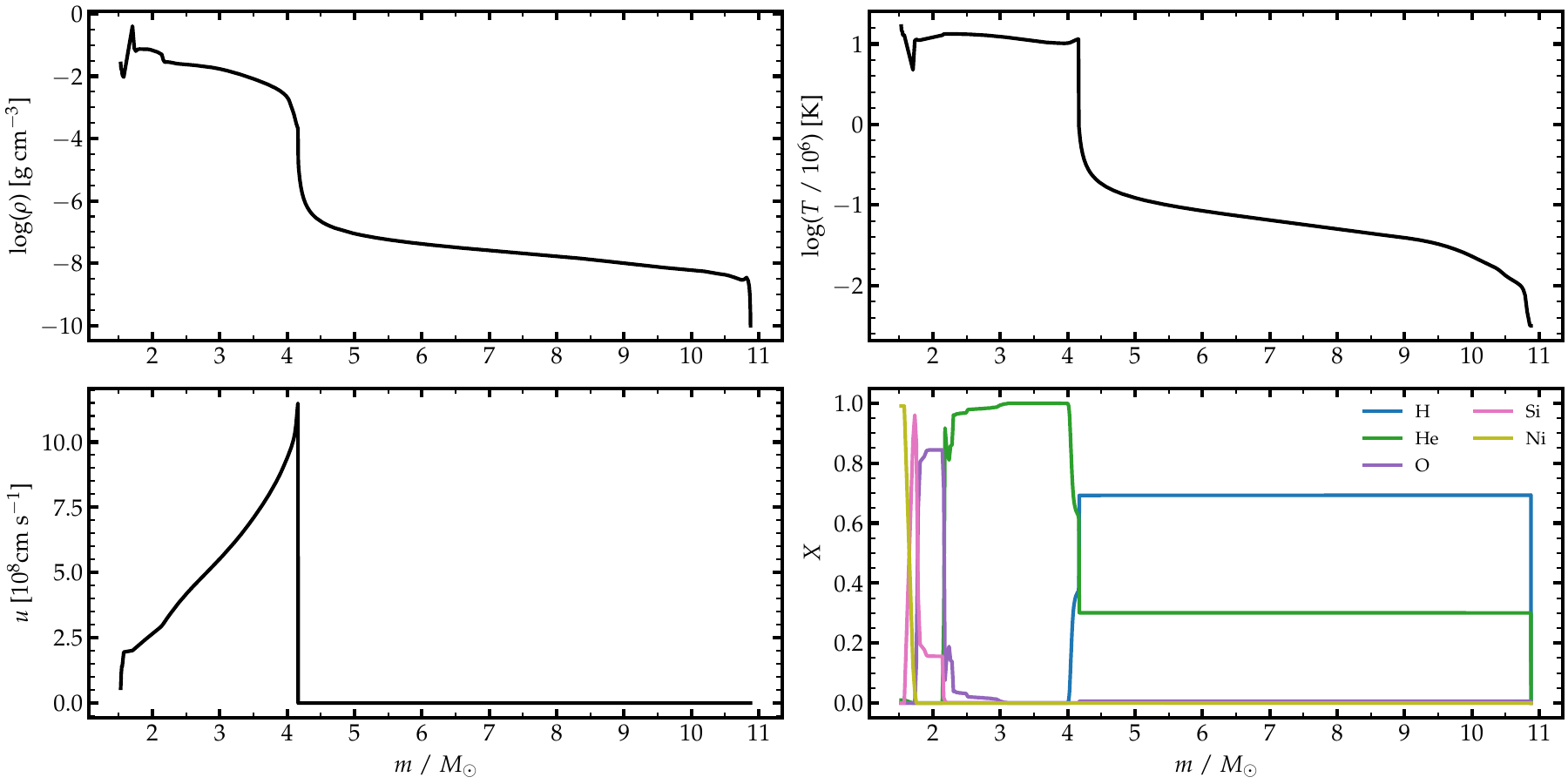}
    \caption{Initial conditions for our \heracles\ simulations. We show the density (top left), temperature (top right), velocity (bottom left), and the mass fraction of H, He, O, Si, and \nickel Ni, as a function of Lagrangian mass. The time is 500\,s after the explosion trigger.}
    \label{fig:init_physics}
\end{figure*}

\subsection{Preamble}
\label{subsec:preamble}

We now turn to the more astrophysics-oriented context of a spherical explosion occurring in a RSG star. This initial work is used here for the benchmarking of the code but it fits within a larger project of simulating the various instabilities taking place in the extended envelopes of RSG stars before, during, and after their explosion. These so-called type II SNe are the most frequently encountered type of SN from massive star explosions (see, e.g., \citealt{Perley_2020}). Furthermore, detailed NLTE radiative-transfer calculations have demonstrated that both small and large-scale structures of SN ejecta have a significant impact on SN light curves and spectra \citep{Dessart_2018, Dessart_2019, Ergon_2022} so progress in that sector is much needed in order to improve the physical consistency of radiative-transfer and radiation-hydrodynamics simulations and the robustness of their inferences.

For this work, we used the initial ejecta conditions of \citet{Dessart2024}, which are based on a subset of evolutionary calculations for binary massive stars performed by \citet{Ercolino2024}. Binarity, which occurs with a high frequency in massive stars \citep{Sana2012}, can lead to a wide range of H-rich envelope masses, $M$(H-env), in the pre-SN star because of mass transfer through Roche-lobe overflow. However, as long as a residual H-rich envelope remains, the He-core mass as well as the surface radii are largely unaffected by this mass loss, so that these objects should explode with similar properties (explosion energy and nucleosynthesis), merely modulated by the differing $M$(H-env). \citet{Dessart2024} indeed showed that through variations in $M$(H-env) alone, the full diversity of type II SN light curves may be reproduced. Chemical mixing and clumping may, however, differ widely between these ejecta due to the differences in their progenitor-envelope properties, with strong implications for their observables. Investigating this diversity will be the subject of a forthcoming study.

We limited ourselves to the model logP3p45 (this stands for the log of the initial system orbital period of 2818\,d) of \citet{Dessart2024}, whose initial orbit was so wide that it entirely avoided binary mass transfer and essentially evolved as a single star. The initial, primary mass is 12.8\,\msun\ and the system's original mass ratio is 0.95. The primary dies with a mass of 10.84\,\msun, a He-core mass of 3.98\,\msun, a small CO core of only about 2.2\,\msun, and a surface radius of 898\,\rsun. The model also exhibits a stratification in composition typical of massive, RSG stars at death. The envelope shows a succession of layers of distinct composition with a massive H-rich envelope of about $\sim$\,6.9\,\msun\ and a degenerate core of $\sim$\,1.5\,\msun. 

In practice, we started our simulations from the output of the explosion simulation carried out with the 1D, Lagrangian radiation-hydrodynamics code \vund\ \citep{Livne1993, Dessart_2010b, Dessart_2010a}, at a time of 500\,s relative to the onset of the explosion trigger. The total energy injected in the star for the explosion was designed to yield a kinetic energy at infinity of $1.1\times10^{51}$\,erg so by this time of 500\,s, the shock has already reached the outer layers of the He core\footnote{For details about the preSN evolution and explosion simulation, we refer the reader to \citet{Dessart2024} and \citet{Ercolino2024}.}. Using \heracles, we first evolved this structure in 1D (Sect.~\ref{subsec:comparison_v1d}) and then in 3D (Sect.~\ref{subsec:3D_results}) in order to study the development of Rayleigh-Taylor instabilities and their impact on chemical mixing and clumping.

\begin{figure}
    \centering
    \includegraphics[width=\columnwidth]{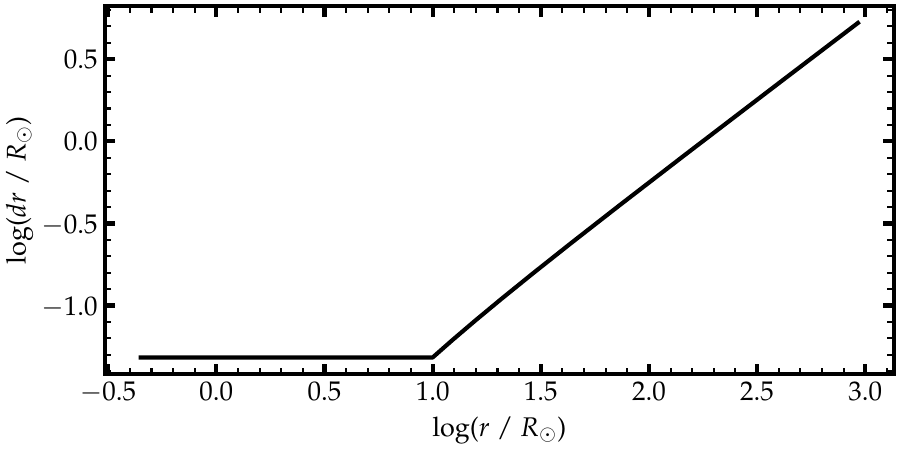}
    \caption{Variation of the radial grid spacing, $dr$, as the function of the radius, $r$. The grid used in the simulation with \heracles\ is regular until 10 \rsun\ and logarithmic beyond.}
    \label{fig:grid}
\end{figure}

%----------------------------------------------------------------------
% Initial setup: profile, grid
%----------------------------------------------------------------------

 \begin{figure*}
  \centering
  \includegraphics[width=\textwidth]{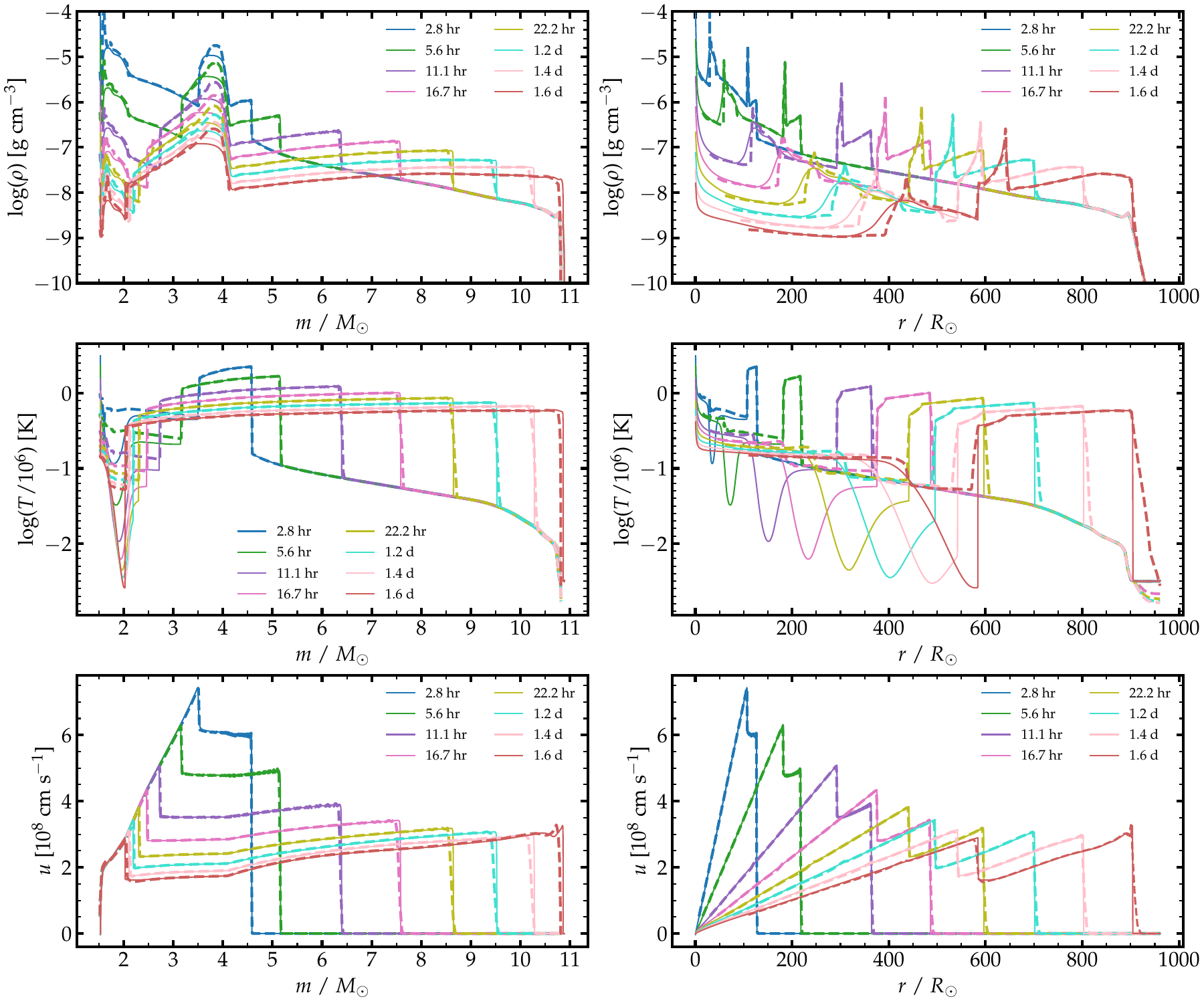}
  \caption{Ejecta properties for our RSG star explosion model as computed by \heracles\ (solid) and \vund\ (dashed). For representative epochs between 2.8\,hr and 1.6\,d, we show the density (top row), the temperature (middle row), and the velocity (bottom row) vs. the Lagrangian mass (left column) and radius (right column). (See Sect. \ref{subsec:comparison_v1d} for discussion.)}
  \label{fig:comparaison_V1D}
\end{figure*}

\subsection{Initial conditions and numerical setup}
\label{subsec:initial_profile}

% initiale profile
The ejecta properties used as initial conditions for the \heracles\ simulations are shown in Fig.~\ref{fig:init_physics}. The shock is located at $\sim$\,4.1\,\msun\ with a peak velocity around 12\,000\,\kms. Beyond the shock, the velocity is essentially zero and $\rho$ and $T$ slowly decrease until the progenitor surface at 10.84\,\msun. Figure~\ref{fig:init_physics} also shows the mass fraction of the main elements (i.e., H, He, O, Si, and \nifs). The position of each layer in the star reflects previous episodes of nuclear burning or mixing, as well as the ashes of explosive nucleosynthesis in the innermost ejecta layers. The extended H-rich envelope is essentially unchanged from its original composition (i.e., corresponding here to solar metallicity), whereas the He-core layers exhibit a stratification of ashes from H, He, C, O, Ne, and Si burning. The explosion energy at 500\,s is primarily stored as kinetic energy, with around $9\times10^{50}$ erg.

Since our interest is to investigate the impact of fluid instabilities on the distribution of elements, we introduced five passive scalars to track the material originally present in the H-rich envelope and in the He-, O-, Si-, or \nifs-rich shells. In the following sections, when referring to the distribution of O, we really mean the distribution of the material in the O-rich shell (in practice, this includes the O/Si, O/Ne/Mg, and the O/C shells).

\begin{figure}
    \centering
    \includegraphics[width=\columnwidth]{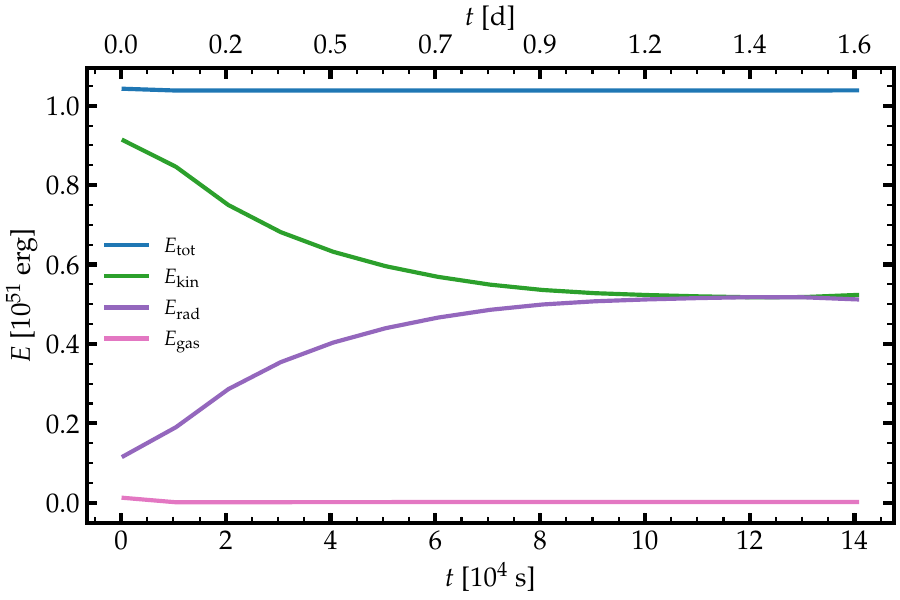}
    \caption{Evolution of the total energy, $E_{\rm tot}$, the kinetic energy, $E_{\rm kin}$, the radiative energy, $E_{\rm rad}$, and the internal (gas) energy, $E_{\rm gas}$, during our \heracles\ simulation from 500\,s until shock breakout. At shock breakout, $E_{\rm kin} \sim E_{\rm rad}$. }
    \label{fig:energy_time}
\end{figure}

% grid
At the start of the \heracles\ simulation, we remapped the \vund\ input (which is on a Lagrangian grid) onto the \heracles\ grid, which we chose to be a hybrid linear-log grid. It has a relatively high resolution out to the shock location (in order to preserve the initial sharpness of the shock and the total initial energy of the explosion) and a decreasing resolution beyond. In Sect.~\ref{subsec:comparison_v1d}, we constructed a regular radial grid with 10000 points to produce a grid that presents no major resolution offset with the V1D grid, which is Lagrangian and thus in mass space (i.e., the goal in this test is to confront the numerical
solution of the hydrodynamical equations between the two codes, rather than assess the influence of resolution or numerical diffusion etc). In Sect.~\ref{subsec:3D_results}, we created a radial grid with a total of 1024 points, with 200 points for the regular part out to 10 \rsun\ (corresponding to a Lagrangian mass of $\sim$\,4.2\,\msun). Figure~\ref{fig:grid} illustrates the variation in the radial increment, $dr$, with radius. In Appendix~\ref{section:impact_resolution}, we consider additional radial resolutions (i.e., $n_r$ of 256, 512, and 768) as part of a resolution study.

% physical setup
We performed hydrodynamical simulations with \heracles, using spherical coordinates, first in 1D (Sect.~\ref{subsec:comparison_v1d}) and then in 3D (Sect.~\ref{subsec:3D_results}). We used the perfect gas+radiation EoS (Eq.~\ref{eq:eos_gp_rad}; we used $\gamma=5/3$ and $\mu=0.62$ for the ideal-gas part) and introduced \nifs-decay heating (Eq.~\ref{eq:heating_ni}; this was done for consistency with \vund\ but it plays little role here since we stopped the simulation at about 1.5\,d, which is only a quarter of the \nifs\ lifetime). A point mass gravity (Eq.~\ref{eq:point_gravity}) was used to account for the presence of the compact remnant. We used a routine to track precisely the internal energy (Sect. \ref{subsec:pressure_fixe}). The boundary conditions were reflexive in the inner radius and transmissive at the outer radius. The slope limiter used was Minmod and the Riemann solver was HLLC. For consistency, this general setup was the same for all similar simulations.

%----------------------------------------------------------------------
% Comparison with V1D
%----------------------------------------------------------------------

 \begin{figure}
    \centering
    \includegraphics[width=\columnwidth]{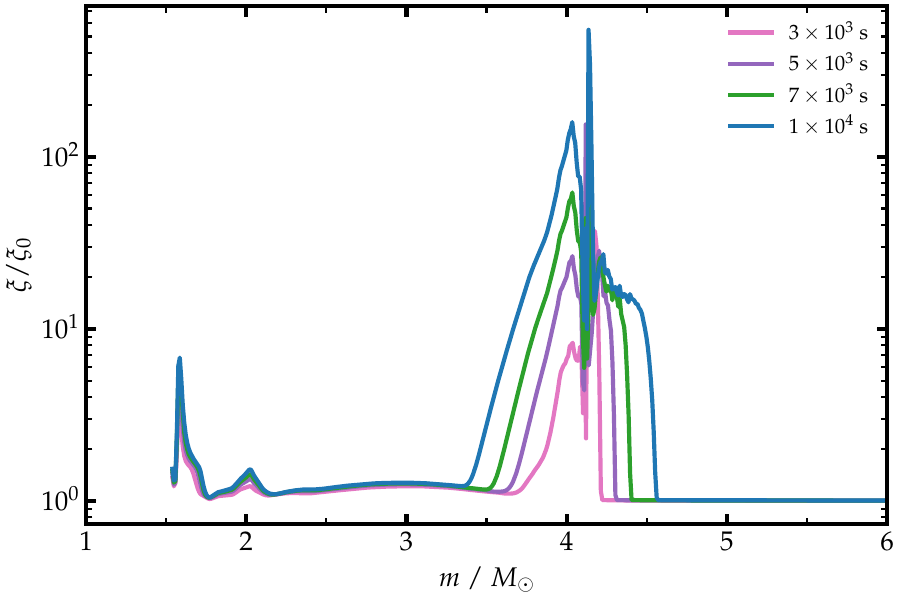}
    \caption{Estimated growth rate of the R-T instability in our RSG explosion simulation, shown at multiple epochs and vs. Lagrangian mass. (See Sect. \ref{subsubsec:stabiliyt_analysis} for discussion.)}
    \label{fig:stab_compressible}
\end{figure}

 \subsection{Results in spherical symmetry and comparison with \vund}
 \label{subsec:comparison_v1d}

 In this section, we first present a comparison of the results between \vund\ and \heracles, and thus in spherical symmetry. Figure~\ref{fig:comparaison_V1D} shows the evolution of $\rho$, $T$, and $u$ over time as the shock propagates through the progenitor envelope and until its emergence at $R_\star$ at $\sim$\,1.6\,d.\footnote{For this figure and to ensure the Eulerian grid is as fine in the inner regions as the Lagrangian grid in \vund, the \heracles\ simulation employs 10\,000 radial zones.} We show the variations in these quantities as a function of both the Lagrangian mass and the radius. There is a good agreement between the results of the two codes.

A pair of reverse and forward shocks forms when the SN shock reaches the outer edge of the He core very soon after the start of the simulation. The reverse shock then propagates slowly inward (in mass space) and will eventually reach the innermost ejecta layers after about 10\,d (as is indicated in the \vund\ simulation; not shown here). The reverse shock plows through the extended envelope more swiftly but it undergoes a strong deceleration because of the relatively flat density profile of the massive extended H-rich envelope. Over the course of the simulation, the ejecta density drops from $10^{-4}$ to $10^{-7}$\,g\,cm$^{-3}$, primarily through the expansion of the dense inner regions over a timescale of days. In contrast, the outer regions, located at large distances from the explosion site, are shocked last and undergo limited expansion. A high-density plateau forms around the peak density so that most of the shocked-envelope mass is located between 500 and 900\,\rsun\ at shock emergence. A small and decreasing mass is present below the reverse shock owing to the large progenitor envelope mass relative to the He-core mass.

The velocity profile mirrors that of the density. From the time of the explosion, the peak velocity decreases from 7500 to 3000\,\kms. Over time, the shock front remains steep while propagating through the envelope. The main difference between the two codes is the prediction of a temperature drop interior to the reverse shock. This feature is also present in \mesa\ but absent in \vund\ because of the use of a temperature (or internal energy) floor in the EoS (for a similar pattern, see Fig.~28 of \citealt{Paxton2015}). There is a good agreement between the two codes for the temperature in the shocked regions, which are radiation-dominated (i.e., $P_{\rm r} \gg P_{\rm g}$).

The deceleration of the forward shock is also reflected by the decrease in the total kinetic energy over time. Figure~\ref{fig:energy_time} shows the evolution of the various energy components versus time from the start of the simulation at 500\,s until shock breakout (for each energy component, the summation is over the full grid).
The total energy is primarily kinetic initially, but is transformed into internal energy (i.e., primarily radiation) as the shock propagates. At shock breakout, the total energy is split essentially equally between radiative energy and kinetic energy. This radiation-dominated shock is at the origin of the burst of radiation taking place at shock emergence, which is not modeled in this hydrodynamics-only simulation.

%----------------------------------------------------------------------

\begin{figure*}
    \centering
    \begin{subfigure}[b]{\columnwidth}
    \centering
    \includegraphics[width=0.9\columnwidth]{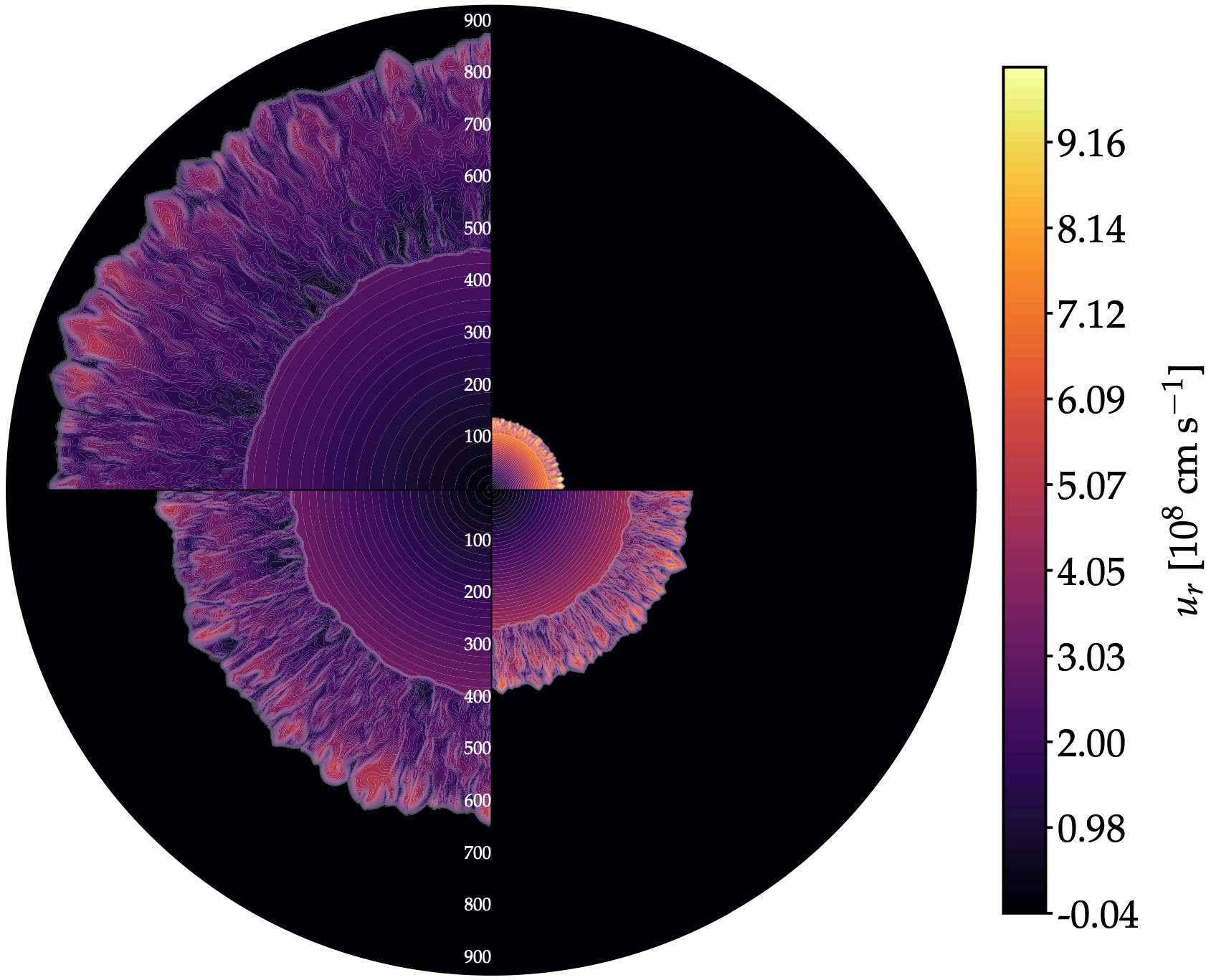}
    \caption{ }
    \end{subfigure}
    \hfill
    \begin{subfigure}[b]{\columnwidth}
    \centering
    \includegraphics[width=0.9\columnwidth]{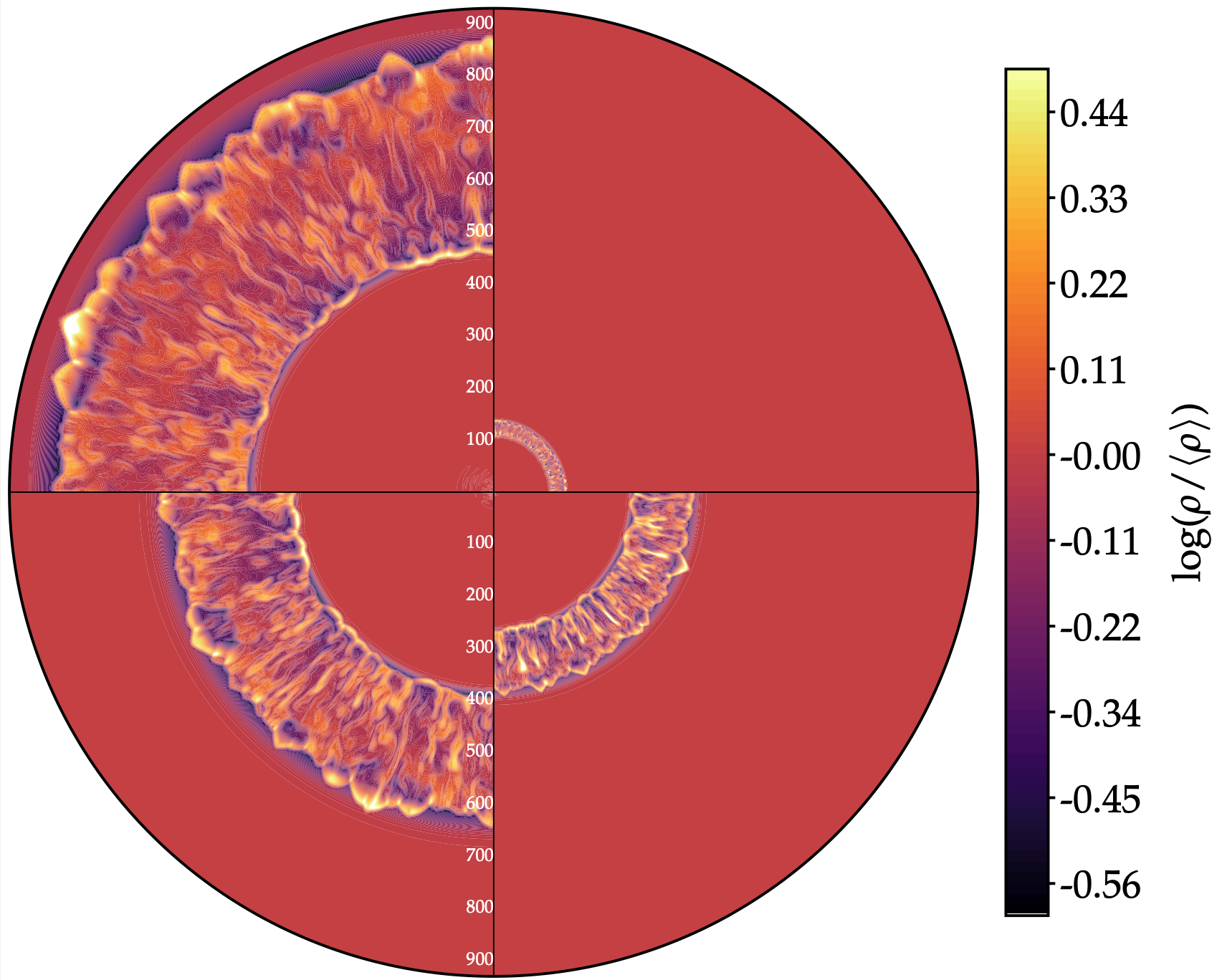}
    \caption{ }
    \end{subfigure}
    \vspace{0.5cm}
    \begin{subfigure}[b]{\columnwidth}
    \centering
    \includegraphics[width=0.9\columnwidth]{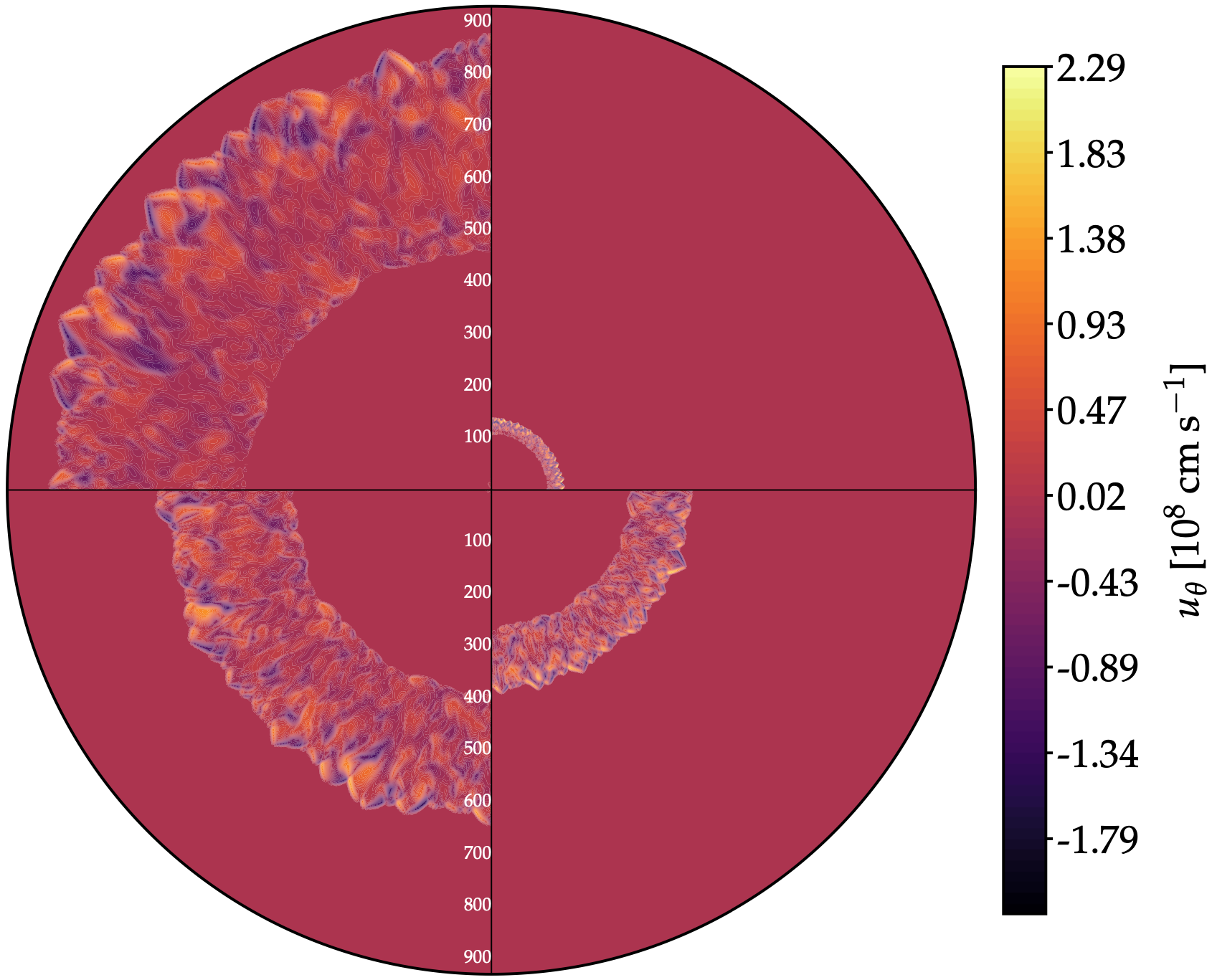}
    \caption{ }
    \end{subfigure}
    \hfill
    \begin{subfigure}[b]{\columnwidth}
    \centering
    \includegraphics[width=0.9\columnwidth]{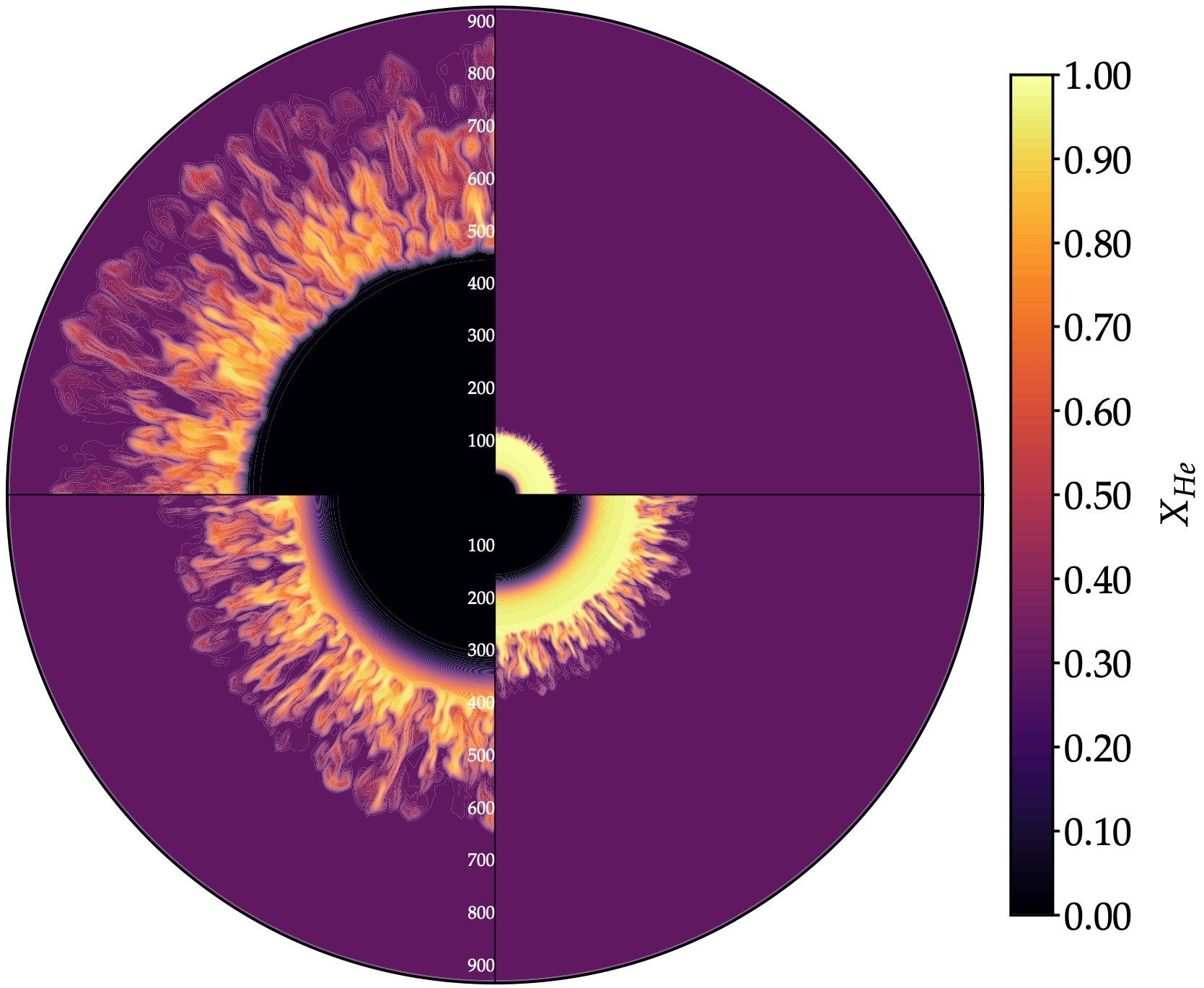}
    \caption{ }
    \end{subfigure}
    \caption{Multiepoch meridional slices for the 3D RSG-star explosion simulation with \heracles. The panels show the radial velocity (a), the normalized density, $\rho/\langle\rho\rangle$ (b), the polar velocity, $u_\theta$ (c), and the He mass fraction, $X_{\rm He}$ (d). Within each panel, each quadrant corresponds to a cut at $\phi=\pi/2$ and progresses clockwise from 1, to 4, 8, and $12\times10^4$\,s. (See Sect. \ref{sect:res_3d} for discussion.)}
    \label{fig:2D_plot}
\end{figure*}

%----------------------------------------------------------------------
% Stability Analysis
%----------------------------------------------------------------------
 
\subsection{Stability analysis}
\label{subsubsec:stabiliyt_analysis}

Before discussing the 3D simulations of our explosion model, we conduct an approximate local stability analysis using the method of \citet{Benz1990} (see also \citet{Muller1991}). It estimates the growth rate of the Rayleigh-Taylor instability in the compressible case as:
\begin{equation}
  \sigma = \frac{c_s}{\gamma}\sqrt{\mathscr{P}^2 - \gamma \mathscr{P} \mathscr{R}} \, ,
\end{equation}
with the following definitions:
\begin{equation}
    \mathscr{R}=\frac{\partial \ln \rho}{\partial r} \quad \textrm{and} \quad  \mathscr{P}=\frac{\partial \ln P}{\partial r} \, ,
\end{equation}
where $c_s$ is the adiabatic sound speed accounting for both the gas and the radiation (Eq.~\ref{eq:cs_gp_rad}). Regions with the largest growth rate with be the most unstable. This estimate is only indicative since it applies to the linear regime of the instability. 

Similarly, the total time-integrated growth at a given mass is given by
\begin{equation}
    \frac{\xi}{\xi_0} = \exp (\int_{t_0}^t \sigma dt)  \, ,
\end{equation}
where $\xi$ ($\xi_0$) is the amplitude of the instability at time $t$ ($t_0$). 

Figure~\ref{fig:stab_compressible} shows the integrated growth rate as a function of the Lagrangian mass at selected epochs (in the incompressible case, $\xi/ \xi_0$ is qualitatively similar but peaks at smaller values). As is expected, the most unstable region is located at the interface between the highly bound He-core and the loosely bound H-rich envelope, at $\sim$\,4\,\msun, and thus where there is a density cliff. The small peak at $\sim$\,1.5 \msun\ is an artifact of the initial ejecta density structure reflecting the regions of the original thermal bomb. This analysis indicates that the Rayleigh-Taylor instability should develop most strongly at the H-He interface in 3D simulations, as was previously found (see, for example, \citealt{Muller1991}).

%----------------------------------------------------------------------
% Simulations 3D
%----------------------------------------------------------------------

\subsection{Results from the 3D simulation}
\label{subsec:3D_results}

%----------------------------------------------------------------------
% Initial perturbations
%----------------------------------------------------------------------
 
\subsubsection{Setup}
\label{subsuec:perturb_init}

The 3D simulation with \heracles\ was initialized with the same 1D input described in Sect.~\ref{subsec:initial_profile}. This 1D profile was remapped onto the same radial grid and copied across all angular directions $(\theta, \phi)$ both in the range $[\pi /4, 3\pi/4]$. We used the same setup as in Sect.~\ref{subsec:initial_profile} and added reflexive boundary conditions for the angular directions. We chose a grid with $[n_r, n_\theta, n_\phi] = [1024,512,512]$. Simulations using other resolutions in both radius and angle are presented in Appendix~\ref{section:impact_resolution}. In the preliminary tests for the 3D simulations, we employed a variety of initial perturbations of the density and velocity to act as seeds for the Rayleigh-Taylor instability. We find, however, that the instability grows from numerical noise and is also aided by the small-scale high-frequency oscillations present in the \vund\ input (i.e., in the post-shock region at the highest velocities). So, no initial perturbation of the fluid variables was introduced for the tests presented in this work.

%----------------------------------------------------------------------
% Results 3D simulations
%----------------------------------------------------------------------

\begin{figure}
    \centering
    \includegraphics[width=\columnwidth]{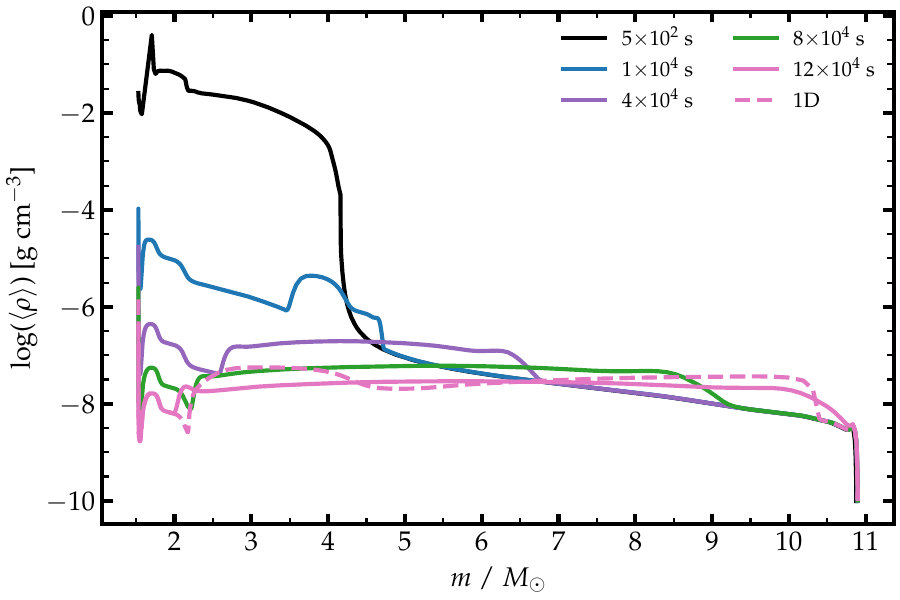}
    \includegraphics[width=\columnwidth]{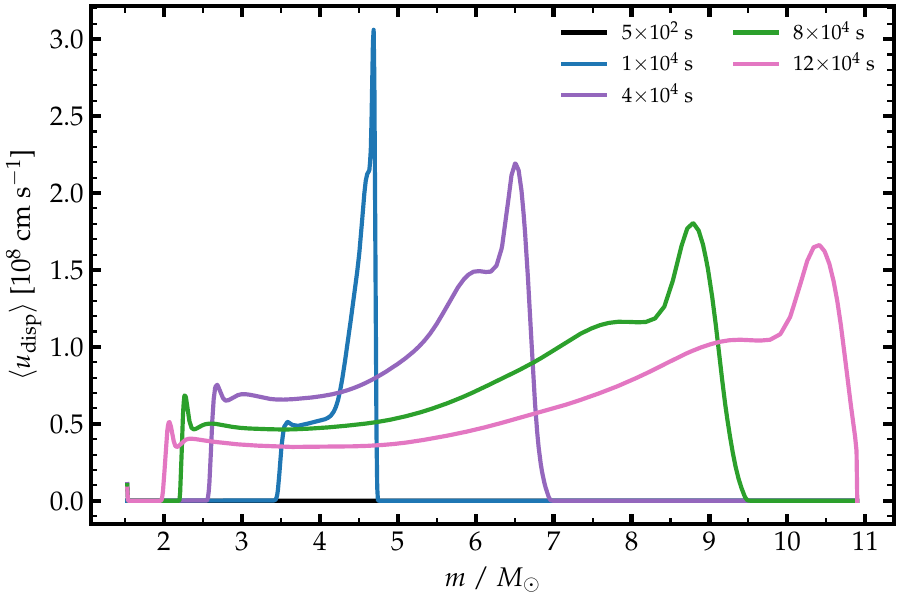}
    \includegraphics[width=\columnwidth]{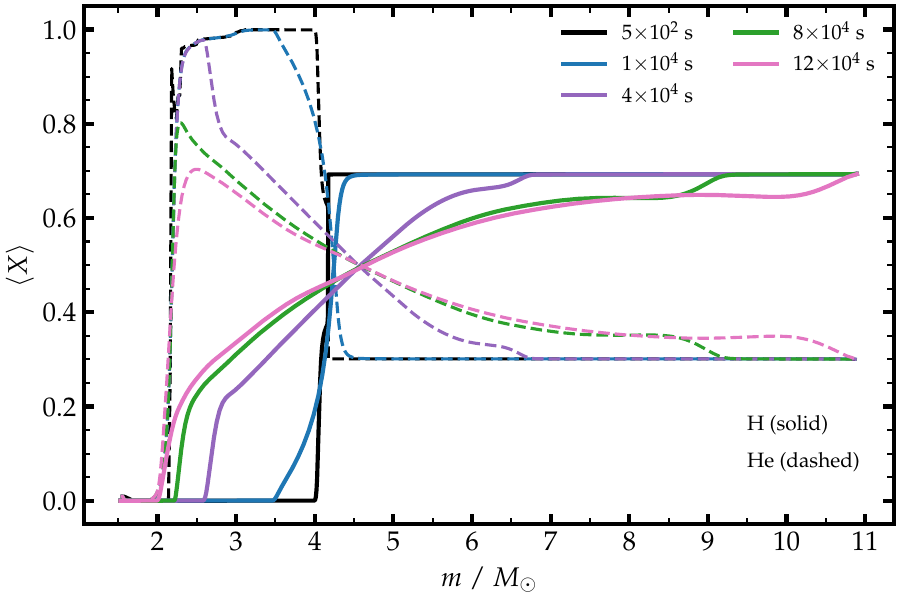}
    \caption{Evolution of the angle-averaged density (top), velocity dispersion (middle), and composition (bottom; the solid [dashed] line corresponds to H [He]) from the initial time until shock emergence and versus Lagrangian mass. We show the same epochs as in Fig.~\ref{fig:2D_plot}. In the top panel, the final density obtained in the 1D counterpart is shown as a dashed line. (See Sect.~\ref{sect:res_3d} for discussion.)}
    \label{fig:3D_time}
\end{figure}

\begin{figure*}
    \centering
    \begin{subfigure}[b]{\textwidth}
    \centering
    \includegraphics[width=\textwidth]{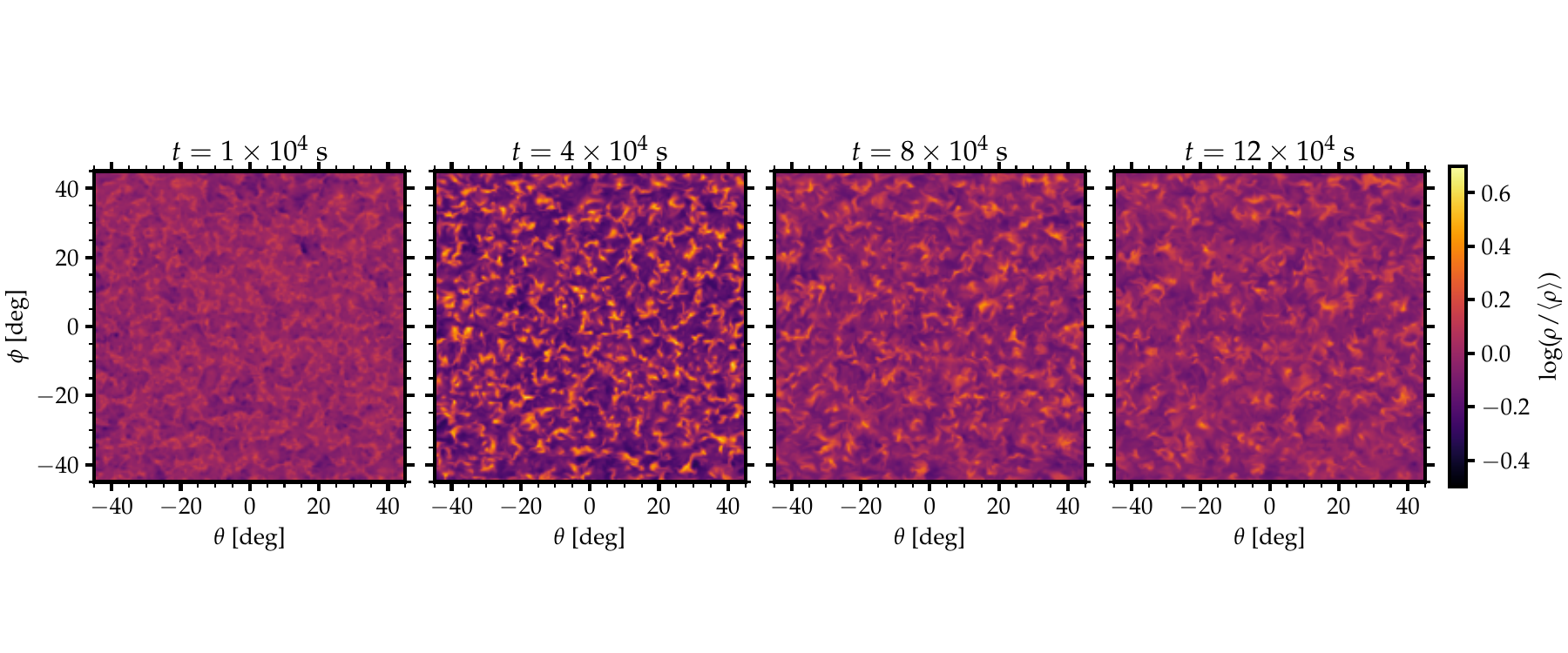}
    \end{subfigure}
    \hfill
    \begin{subfigure}[b]{\textwidth}
    \centering
    \includegraphics[width=\textwidth]{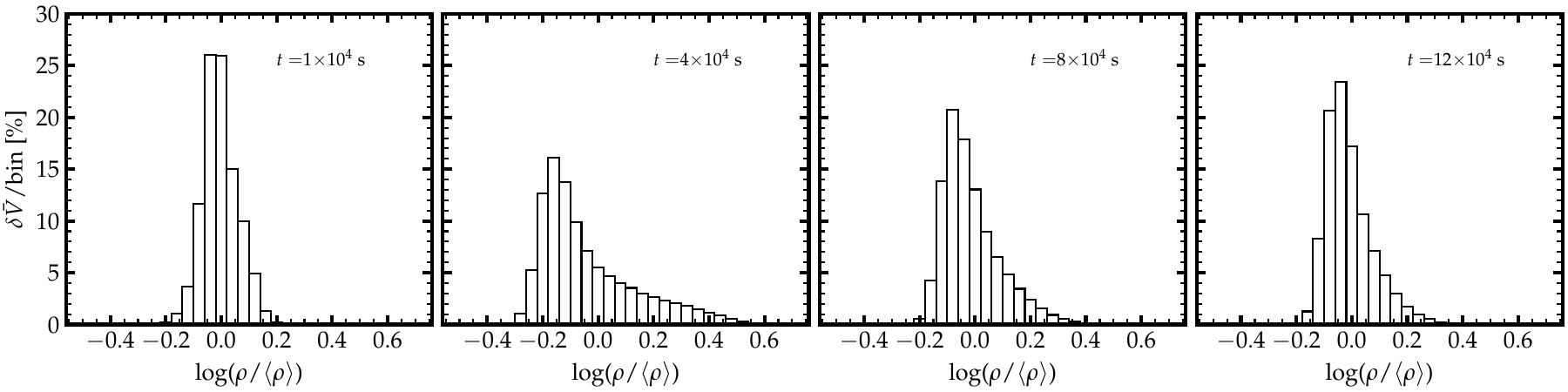}
    \end{subfigure}
    \caption{Evolution of the angular properties of the quantity $\log(\rho/\langle\rho\rangle$) at a fixed Lagrangian mass (4.5\,\msun). In the top row and from left to right, we show color maps of radial sections through the 3D domain at 127, 323, 500, and 635\,\rsun\ and at times of 1, 4, 8, and 12 $\times 10^4$\,s, respectively. In the bottom row, we show the fractional shell volume per normalized density bin for each of the shells shown in the top row.} 
    \label{fig:coupe_th_ph}
\end{figure*}

\begin{figure*}

    \centering
    \includegraphics[width=\textwidth]{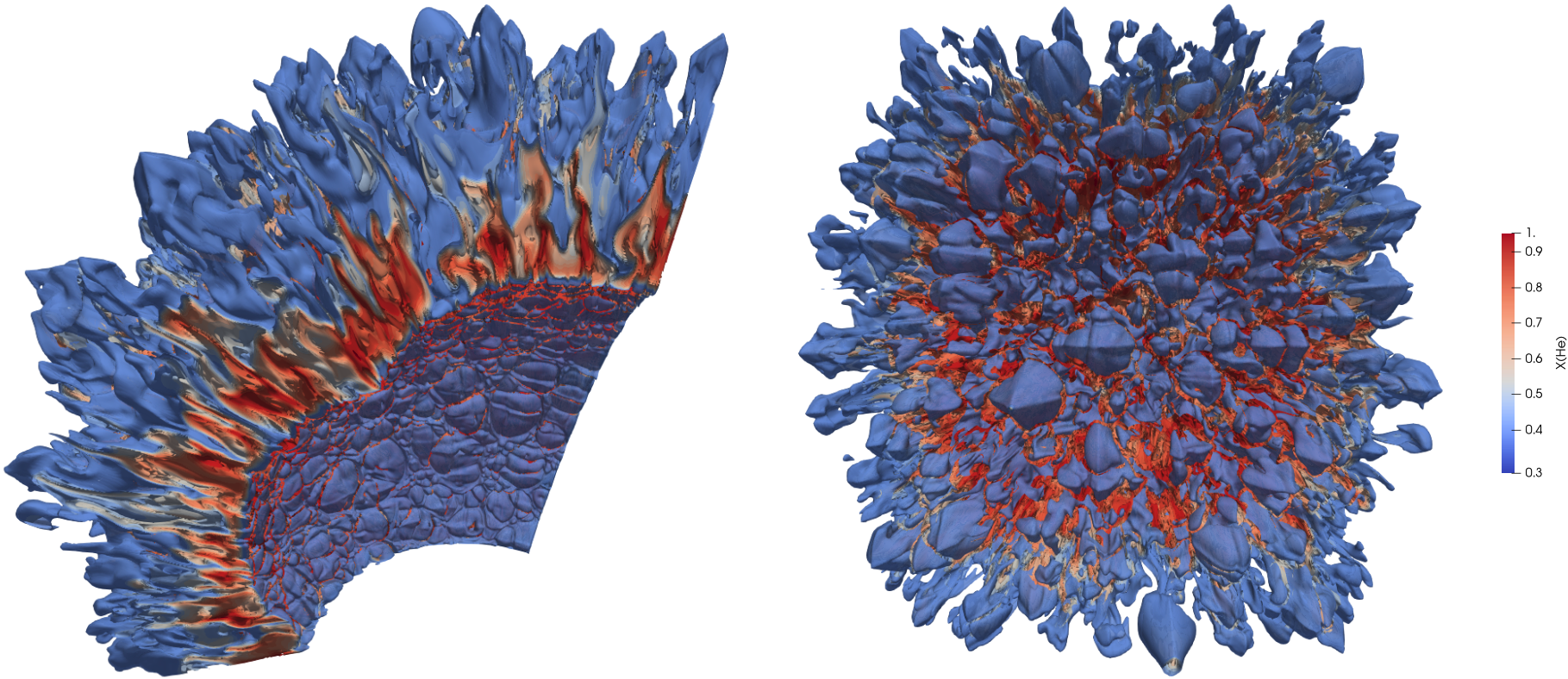}
    \caption{3D rendering of the helium mass fraction in the SN test for two different viewing angles. The time is $12\times10^4$\,s.}
    \label{fig:3D_plot}
\end{figure*}

\subsubsection{Results}
\label{sect:res_3d}

In this section, we present a 3D simulation of the perturbed 1D explosion model presented in Sect.~\ref{subsec:initial_profile}. To facilitate the discussion of the results, we first present the evolution of various fluid quantities within meridional slices through the 3D domain. Then, we also discuss angle-averaged quantities for the same epochs but shown versus the Lagrangian mass rather than the radius. In the following, ``$<>$'' is meant as an average over the polar and azimuthal angles at a given radius or Lagrangian mass. 

Figure~\ref{fig:2D_plot} shows the radial velocity, the logarithm of the normalized density, the polar velocity, and the He mass fraction  in the $\phi = \pi/2$ plane at 1, 4, 8, and $12\times10^4$\,s -- each quadrant corresponds to a time ordered chronologically from twelve noon. These slices are representative of the overall evolution of the 3D ejecta and other choices for these slices (i.e., a different azimuth, or cuts instead in the $(r,\phi)$ plane exhibit similar properties). 

As in 1D, we see that a pair of reverse and forward shocks form soon after the start of the simulation. However, in contrast to the 1D simulation, fluctuations in the velocity, the density, and the composition quickly appear throughout the region in between the shocks. These shocks are clearly visible in Fig.~\ref{fig:2D_plot}a, bracketing a turbulent flow that is located between about 450 and 900\,\rsun\ at $12\times10^{4}$\,s. The reverse shock remains roughly spherically symmetric but the forward shock is corrugated, showing variations in extent at the 5--10\,\% level. The structured, unstable flow extends all the way to the forward shock. This probably arises from the strong deceleration of the forward shock as it propagates through the extended RSG star envelope (compare the first and last quadrants in Fig.~\ref{fig:2D_plot}a; see also Fig.~\ref{fig:comparaison_V1D}).

Figure~\ref{fig:2D_plot}b indicates that the density variations are maximum at the forward shock, reaching a factor of a few the angular average at that radius, and to a lesser extent at the reverse shock. This is partially an artifact from showing the density normalized to the average and the fact that the forward shock is corrugated. Within the shocked region, the density enhancements appear as filaments surrounded by more extended, lower-density regions that are close to the average density.
 
The meridional slice of the polar velocity (Fig.~\ref{fig:2D_plot}c) shows that the horizontal velocity is a sizable fraction of the radial velocity, with maximum values reached at the forward shock. In the turbulent region below the forward shock, the angular velocity is subsonic and lower than about 1000\,\kms\ at all times ($c_s\sim$\,2000\,\kms\ in this turbulent region) -- the angle-averaged polar and azimuthal velocities are essentially zero. 

The morphology of the instability-generated structures is more vividly illustrated by the He distribution, which is the dominant element in the He-rich shell (at the 95\,\% level) and the second-most abundant in the H-rich shell (at the 28\,\% level). In Fig.~\ref{fig:2D_plot}d, we see the development of He-rich fingers, plumes, and mushrooms into the H-rich envelope, stretching out as time passes, and reaching out to the forward shock at the last time shown. These structures are well resolved in our calculations, which use 512 polar and azimuthal zones for full coverage in each direction of ninety degrees (see Appendix~\ref{section:impact_resolution} for a comparison of the structure with different radial and angular resolutions). The He mass fraction in the most radially elongated structures is, however, much smaller than the original He mass fraction in the He-rich shell. These structures have probably expanded laterally and mixed with the H-rich material, which may in part arise from numerical diffusion.

Figure~\ref{fig:3D_time} shows the evolution (at the same epochs as in Fig.~\ref{fig:2D_plot}, apart from the initial time) of the angle-averaged density, velocity dispersion,\footnote{$\langle u_{\rm disp}\rangle\equiv \sqrt{\langle u_r^2\rangle - \langle u_r\rangle^2 + \langle u_{\theta}^2\rangle + \langle u_{\phi}^2\rangle}$.} and composition from the initial time until shock emergence and versus Lagrangian mass. Although the density contrast is strongly reduced because of expansion, we see that in 3D this contrast is further reduced so that the average density is essentially constant throughout the shocked envelope. It no longer shows a jump at the Lagrangian mass corresponding to the outer edge of the He core (see also Fig.~\ref{fig:comparaison_V1D}). High-density extrema are still present at the forward and reverse shocks but they no longer have the $4\pi$ lateral coherence scale enforced in 1D, which tends to produce a softer density variation at the shocks. Both shocks stretch further in and out than what is obtained in the 1D counterpart. The velocity dispersion (middle panel of Fig.~\ref{fig:3D_time}) shows a similar behavior to the polar velocity in the meridional slice shown in Fig.~\ref{fig:2D_plot}, with values that asymptote at shock emergence to 500-1000\,\kms. Perhaps the most striking impact of the instabilities taking place until shock emergence is the strong chemical mixing (bottom panel of Fig.~\ref{fig:3D_time}). In 1D, the original composition profile would remain unchanged when viewed relative to the Lagrangian mass coordinate. In 3D, the composition is instead progressively altered with the mean mass fraction of H increasing inward and the mean mass fraction of He increasing outward. This pollution stretches all the way to the reverse shock for H (into ejecta regions that are H deficient) and to the forward shock for He (into regions that are already He rich). At the end of the simulation at $12\times10^4$\,s, the reverse shock barely reaches the O-rich shell where we expect a secondary Rayleigh-Taylor instability to develop \citep{Muller1991}.

To characterize more quantitatively the lateral structure in the 3D simulation, Fig.~\ref{fig:coupe_th_ph} shows the properties of a slice of the quantity $\log(\rho/\langle\rho\rangle)$ at different radii and times, and starting at 10$^4$\,s. We selected a slice at 120\,\rsun\ initially (around a Lagrangian mass of 4\,\msun), which is located just below the forward shock at 137\,\rsun\ (the choice of the radius at the different times is made such that we track the same slice of the ejecta as it advects outward). The top row shows color maps of the density variations relative to the polar and azimuthal angles, whereas the bottom row shows histograms of the fractional volume (within that slice or shell) per bin in $\log(\rho/\langle\rho\rangle)$. At 10$^4$\,s, structures are within 50\,\% of the average, with naturally a larger volume occupied by the underdense material (the histogram is slightly skewed relative to the mean density). At the next epoch, the density contrast between underdense and overdense material has grown, with the density maximum reaching four times the average density. The histogram is now highly skewed with about 65\,\% of the volume occupied by material about 50\,\% below the average density. The structures are randomly distributed in angle (and remain so at all times). As time progresses and the ejecta expand further, the histogram becomes more symmetric and the entire material is found within 50\,\% of the mean density. There is thus only a modest residual density contrast in the shocked envelope by the time the shock emerges from the progenitor star. 

Finally, we show a 3D rendering of the He mass fraction in Fig.~\ref{fig:3D_plot} for two lines of sight looking radially inward from outside as well as from the side. This gives an alternative view of the random distribution of the advected He-rich structures into the H-rich envelope of the progenitor star as well as the filamentary nature of the most He-rich material advecting out. The low He-mass fraction of the structures located at the largest radii suggests some level of numerical diffusion.

%--------------------------------------------------------------------
%--------------------------------------------------------------------

\section{Conclusion}
\label{section:conclusion}

In this paper, we have presented the new multidimensional hydrodynamics code \heracles, which is designed for computing at exascale, with a parallelization that works both for CPUs and GPUs, and equipped with a good portability across a variety of architectures. The code is written in C++ and makes use of the Kokkos library. \heracles\ is a Eulerian code that works in Cartesian and spherical coordinates and in 1D, 2D, or 3D. Passive scalars are employed to handle the evolution of a nonuniform initial composition. The mesh may be irregular, adopting for example a combination of a linear grid and a logarithmic grid. The code currently handles constant, point-mass, and interior-mass gravity. Two EoSs are possible with either a pure ideal gas or a mixture of ideal gas and radiation. The code accepts a user-supplied heating or cooling term in order to treat astrophysical conditions such as radioactive decay in SN ejecta or radiative losses from the photosphere of a star. The code was benchmarked with a variety of tests. We show here the results for some usual 1D tests (e.g., a 1D Sod shock tube or a 2D implosion) as well as a 2D Cartesian Kelvin-Helmholtz test, a 3D Cartesian Rayleigh-Taylor test, and a 3D Spherical off-center blast wave test. Performance tests (see Appendix~\ref{section:performance}) were also carried out and reveal a satisfactory scalability of the code.

We also conducted a more astrophysics-oriented test with the study of the propagation of a SN shock through the envelope of a RSG star, using as initial conditions a model of an exploding massive star \citep{Ercolino2024,Dessart2024}. This is a useful test, which has been studied in detail in the context of SN\,1987A by \citet{Fryxell1991} and \citet{Muller1991}. In 1D, \heracles\ produces results for the evolution of the density, temperature, or velocity that are in excellent agreement with those obtained with the 1D Lagrangian (radiation) hydrodynamics code \vund\ \citep{Livne1993}. Because this configuration is Rayleigh-Taylor unstable, we repeated this calculation in 3D and evolved the model until the SN shock emerged at the surface of the progenitor star. We find that the instability develops from the earliest times, eventually extending throughout the shocked region, and leading to filamentary structures or fingers, corresponding to variations in density of the order of a few at several 10$^4$\,s and dropping to a few tens of percent by the time of shock emergence. The instability also erases the density jump that persists in 1D counterparts at the base of the H-rich layers (i.e., the interface between the H-rich envelope and the He core in the pre-SN star). There is dispersion in velocity both in the radial and angular directions, which may survive as turbulence in the ejecta after shock emergence. The most vivid illustration of this instability can be seen in the impact it has on the composition, with advection of H-rich gas down to the H-free innermost regions and advection of He-rich gas from the He-rich shell throughout the H-rich envelope. This behavior qualitatively reproduces previous studies on similar shocked envelopes. 

This benchmarking test will serve as a basis of a forthcoming study on the shock propagation in supergiant stars differing in H-rich envelope mass as well as the long-term influence of \nifs-decay heating in such shocked envelopes. Future work will also extend \heracles\ to a full radiation-hydrodynamics code.

The code is available on GitHub. \footnote{https://github.com/Maison-de-la-Simulation/heraclespp}

%--------------------------------------------------------------------
%--------------------------------------------------------------------

\begin{acknowledgements}
This work was supported by the Paris Region. This project was granted access to the HPC resources of CINES under the allocations 2023-cin4698 and 2024-c1615137 made by GENCI.
\end{acknowledgements}

%--------------------------------------------------------------------
%--------------------------------------------------------------------

\bibliographystyle{aa}
\bibliography{biblio}

%--------------------------------------------------------------------

\begin{appendix}

\section{Resolution study}
\label{section:impact_resolution}

\begin{figure}
    \centering
    \includegraphics[width=\columnwidth]{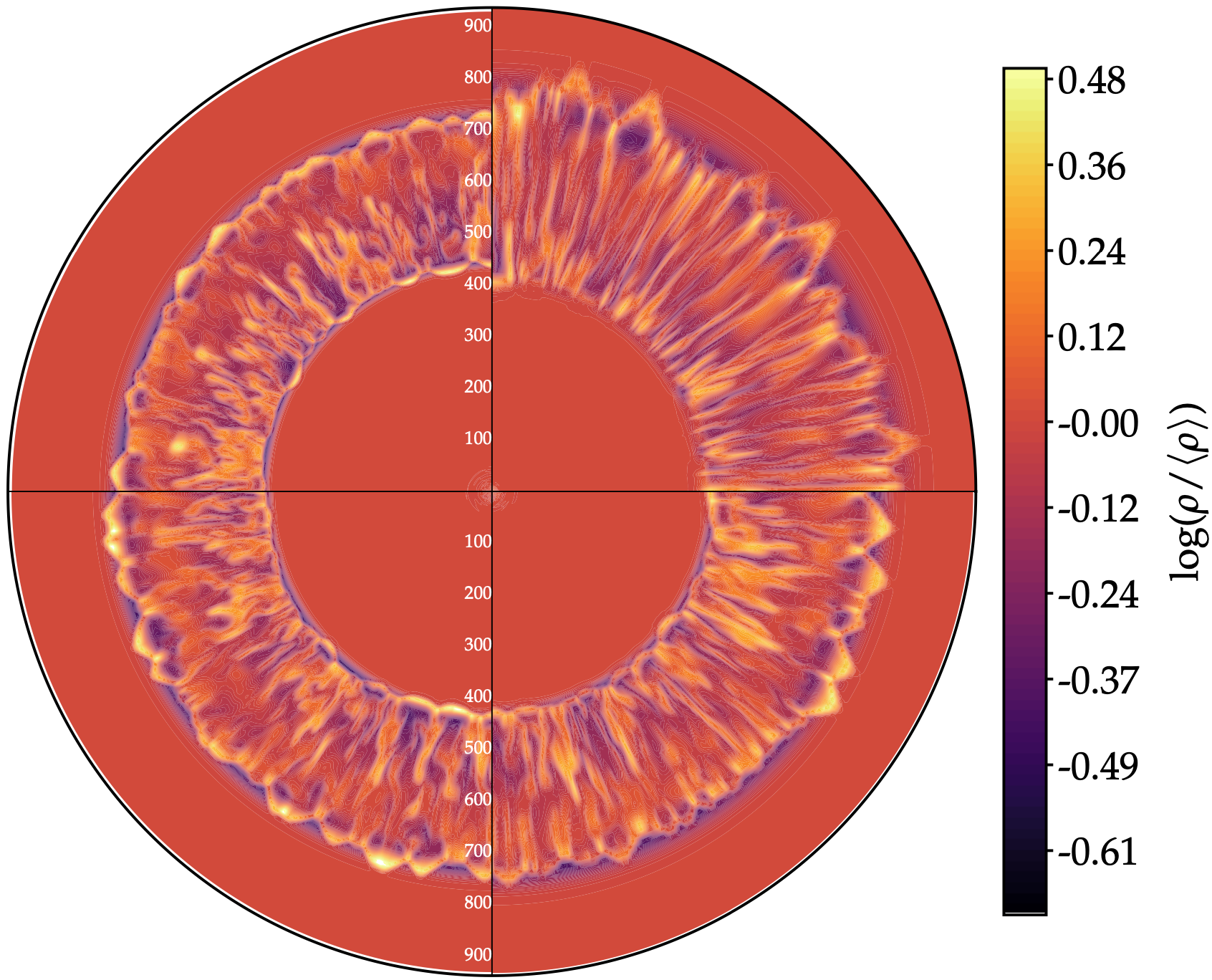}
    \caption{Impact of the radial resolution on the results of the 3D SN shock simulation presented in Sect.~\ref{sect:res_3d}. We show color maps of the quantity $\log(\rho/\langle\rho\rangle)$ in a meridional slice at $\phi=\pi/2$ and corresponding to a time of $10^5$\,s. Clockwise from twelve noon, we show the results for increasing radial resolution by varying $n_r$ from 256 to 512, 768, and 1024 -- the angular resolution is kept fixed with $n_{\theta} = n_{\phi} = 256$.}
    \label{fig:res_rad_3D}
\end{figure}

\begin{figure}
    \centering
    \includegraphics[width=\columnwidth]{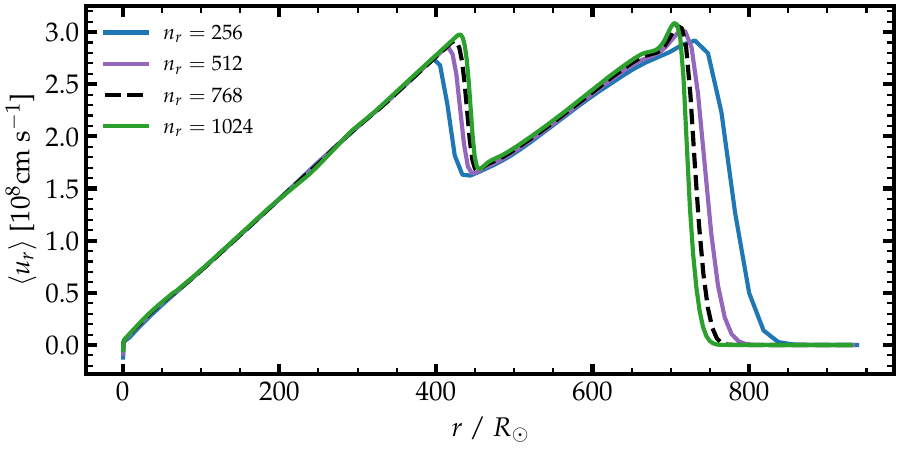}
    \caption{Angular average of the radial velocity versus radius for the simulations that differ in radial resolution and shown in Fig.~\ref{fig:res_rad_3D}.}
    \label{fig:res_rad_1D}
\end{figure}

\begin{figure}
    \centering
    \includegraphics[width=\columnwidth]{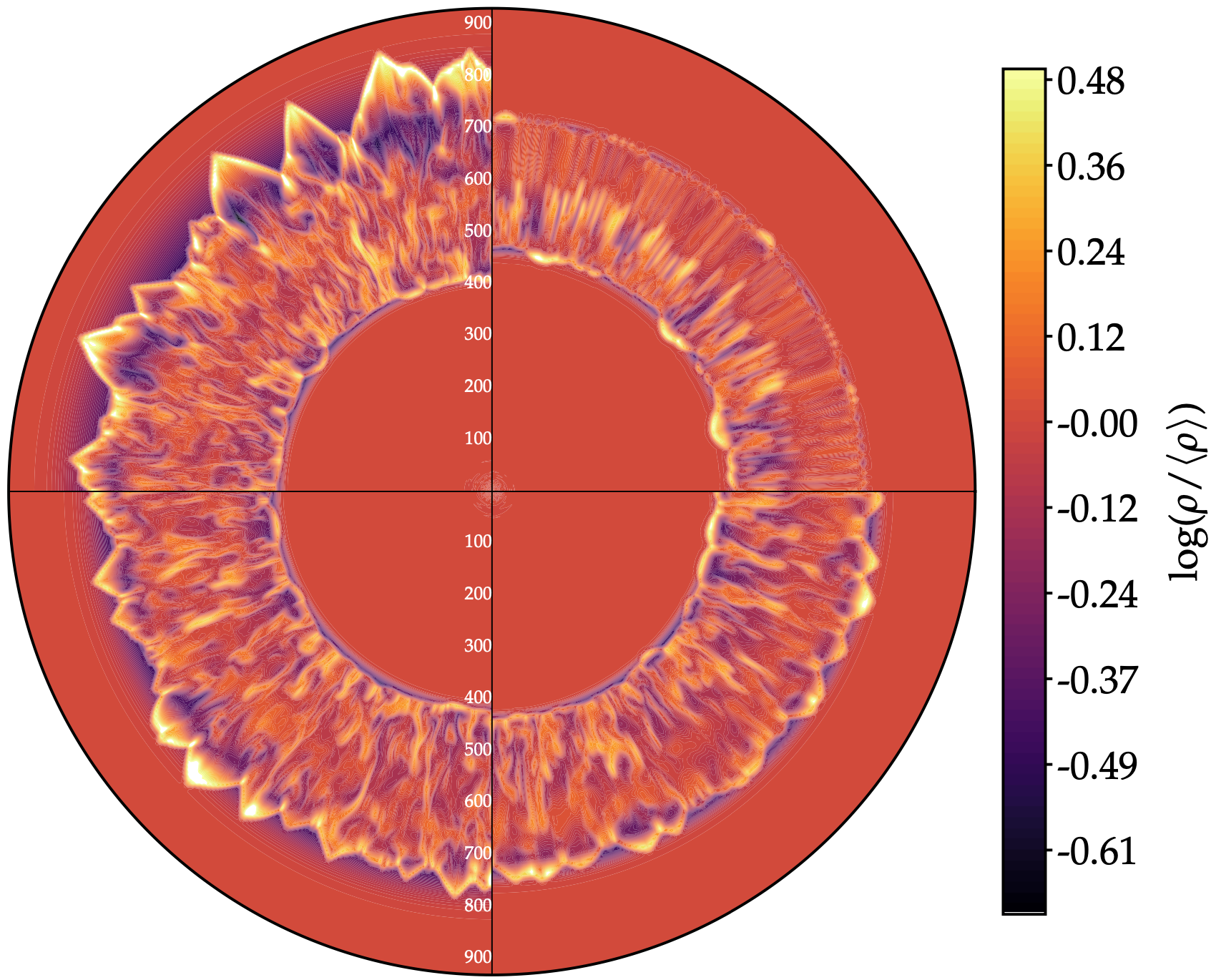}
    \caption{Same as Fig.~\ref{fig:res_rad_3D} but for different angular resolutions. Clockwise from twelve noon, the number of polar and azimuthal zones increases from 128 to 256, 512, and 768 (with $n_{\theta} = n_{\phi}$). The number of radial zones is kept fixed at 768.}
    \label{fig:res_ang_3D}
\end{figure}

\begin{figure}
    \centering
    \includegraphics[width=\columnwidth]{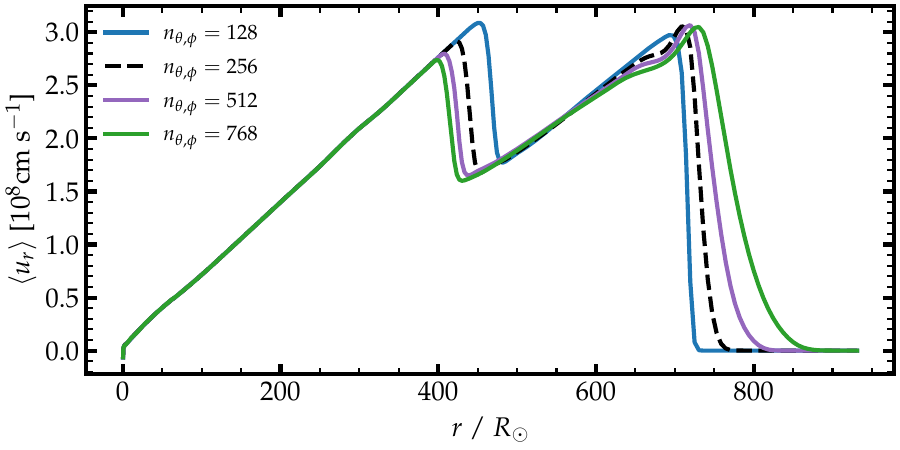}
    \caption{Same as Fig.~\ref{fig:res_rad_1D} but for the simulations differing in angular resolution and shown in Fig.~\ref{fig:res_ang_3D}.}
    \label{fig:res_ang_1D}
\end{figure}

We now present the results from resolution tests. We use the same setup as in Sect.~\ref{subsec:3D_results} but now vary the number of radial zones or the number of angular zones to assess the impact on the R-T instability-generated structures at the time near shock emergence.

Figure~\ref{fig:res_rad_3D} shows meridional slices of the quantity $\log(\rho/\langle\rho\rangle)$ at $10^5$\,s with four simulations in which the number of radial zones was increased from 256,  to 512, 768, and 1024, while the number of angular zones is kept fixed to 256 in both polar and azimuthal directions. Several effects are visible. First, depending on the resolution the initial structure is not equally well resolved so the initial model is resampled into the \heracles\ grid with a resolution-dependent accuracy. At low resolution, both the reserve and forward shocks propagate to smaller and larger radii (see also Fig.~\ref{fig:res_rad_1D}), respectively (for the forward shock, this feature is probably aided by the poorer resolution at larger radii). The post-shock structures are also more radially elongated at lower resolution. However, for $n_r$ of 768 and 1024, the final structures look similar and the overall dynamics (e.g., the position of the forward and reverse shocks) is similar.

Figure~\ref{fig:res_ang_3D} is a counterpart of Fig.~\ref{fig:res_rad_3D} but now showing the results for both $n_\theta$ and $n_\phi$ increased from 128 to 256, 512, and 768, while the number of radial zones is kept fixed at 768. In contrast to the radial-resolution dependence, the models with higher angular resolution produce a pair of reverse and forward shocks that reach smaller and bigger radii, respectively (see also Fig.~\ref{fig:res_ang_1D}. The Rayleigh-Taylor instability-generated structures have a smaller angular scale at higher angular resolution although they are clearly present and to some extent resolved even with the lower resolution employed here. This low resolution corresponds to 0.7\,deg and is already greater than typically employed in 3D simulations of core-collapse SNe (for example, an angular resolution of 1\,deg is used in \citealt{Wongwathanarat2015}).

This study helps us define the size of the simulation for the astrophysical test of the code (Sect.~\ref{section:test_astro}). We needed to balance good resolution of the physics while performing simulations of reasonable size, given the available computational time, resources, and our ability to analyze them. We chose to work with a grid having $n_r=1024$ and $n_\theta=n_\phi=512$, corresponding to a total of about $2.7\times10^8$ cells.

\section{Performance evaluation}
\label{section:performance}

We conduct a weak scaling test to assess the ability of the code to improve its performance when the workload and computing resources are increased in the same proportion. The test is performed on the French supercomputer Adastra\footnote{Adastra Top500: https://www.top500.org/system/180051/} hosted at CINES in Montpellier, France. This cluster is equipped with AMD MI250X GPUs, with four GPUs per node. Each GPU has a 128\,GB memory and a theoretical peak memory bandwith of 3276\,GB/s. The MI250X GPU is composed of two Graphic Complex Die (GCD) and we assign one MPI process to each GCD. We also use the 5.5.1 version of rocm with the HIPCC compiler driver. The Kokkos version is 4.4.01 and we enable the Kokkos option \verb|Kokkos_ARCH_VEGA90A|. The performance is evaluated for the temporal loop without the initialization and I/O. It shows the number of cells updated per second given by $\textrm{perf} = (n_x\, n_y \, n_z \, n_{\rm itr}) / t_{\rm elaps}$, with $n_{\rm itr}$ the number of iterations and $t_{\rm elaps}$ the elapsed time for the calculation.  

\begin{figure}
    \centering
    \includegraphics[width=\columnwidth]{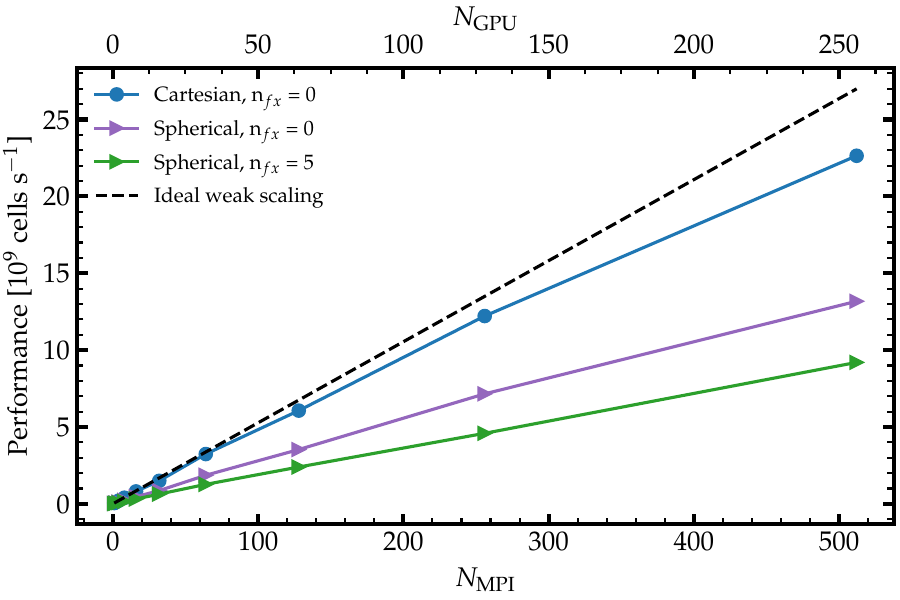}
    \caption{Weak scaling test showing the performance of the code when the size of the problem and the number of MPI processes are increased in the same proportion. We show this performance for different configurations (Cartesian vs spherical coordinates; with and without passive scalars).}
    \label{fig:weak_scaling}
\end{figure}

For this analysis, we used the 3D Rayleigh-Taylor test in Cartesian coordinates discussed in Sect.~\ref{section:RT_3D}. Initially, we do not employ any passive scalar. We maximize the size of the problem with one GCD, which is $5\times 10^7$ cells. After determining the maximum size on a single GPU, we repeated the calculation, each time doubling both the size of the problem and the number of GPUs. The results are shown in Fig.~\ref{fig:weak_scaling}. The ideal weak scaling is calculated based on the performance of a single MPI process and is represented as a line from this assumption. Our results slightly differ from this line.

We then compared with a test based on the configuration for the  propagation of a SN shock in a stellar envelope and discussed in Sect.~\ref{section:test_astro}. The performance is now significantly lower with $1.3\times10^{10}$\,cells\,s$^{-1}$ compared to $2.3\times 10^{10}$\,cells\,s$^{-1}$ in the Cartesian case and 512 MPI processes. This decrease is due to the additional geometrical terms appearing in the equations when spherical coordinates are used as well as the use of an EoS with a mixture of ideal gas and radiation. The use of passive scalars also comes with a sizable computational cost, further reducing the performance in \heracles.

We have also tested \heracles\, on a variety of other architectures including Nvidia GPU's from the IDRIS computing center and various kinds of CPUs (from standard notebook to supercomputers). Thanks to the usage of Kokkos, \heracles\, can run efficiently on all these systems just by changing the compiling options.

\section{Additional hydrodynamic tests}
\label{section:test_hydro_appendix}

\begin{figure}
    \centering
    \includegraphics[width=\columnwidth]{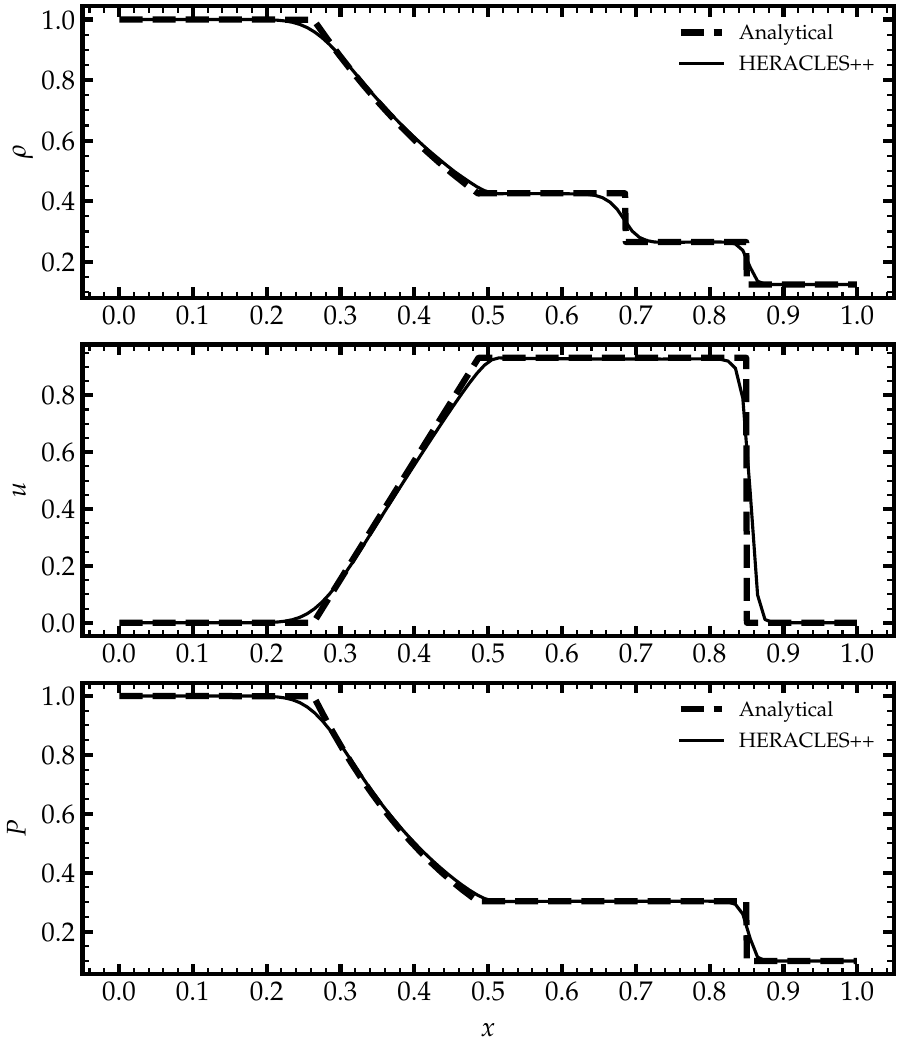}
    \caption{Results for the 1D Sod shock tube test. We show the profiles at $t=0.2$ for the density (top), the velocity (middle), and the pressure (bottom) obtained with \heracles\ (solid) and predicted analytically (dashed).}
    \label{fig:tube}
\end{figure}

\begin{figure*}
    \centering
    \includegraphics[width=\textwidth]{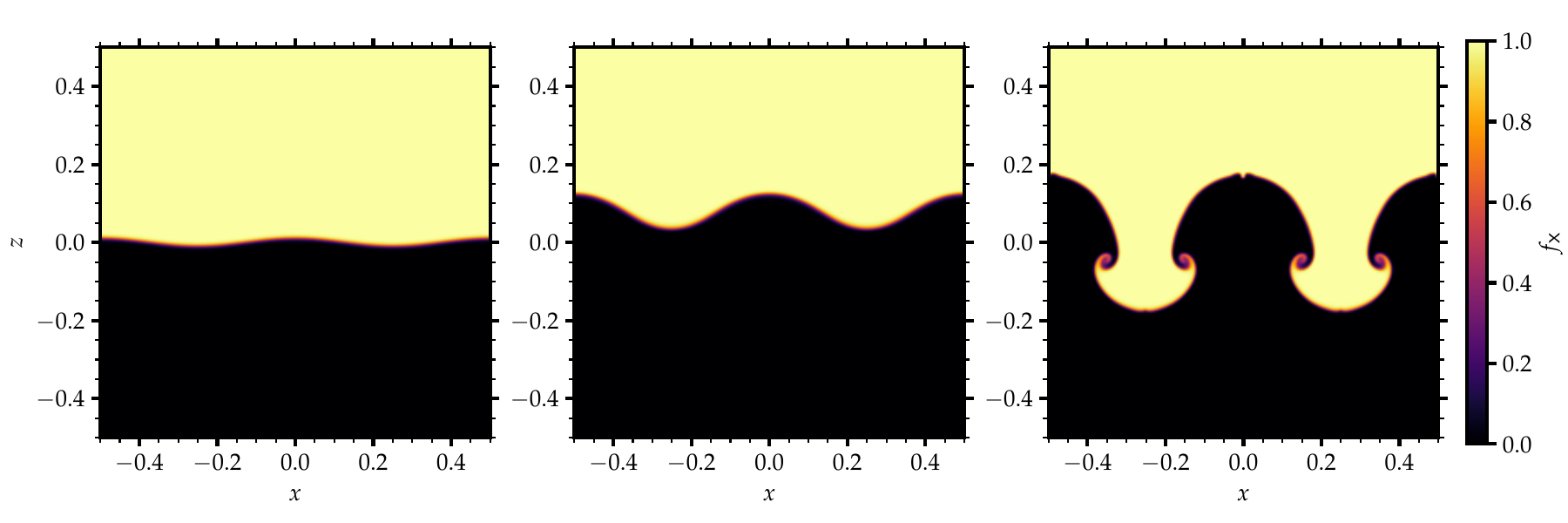}
    \caption{Evolution of the passive scalar $f_{\rm x}$ for the 2D Rayleigh-Taylor instability test. From left to right, we show the results at $t=0$, 1.5, and 3. (see Appendix~\ref{sect_appendix:2d_rt_test} for details.)}
    \label{fig:RT_2D}
\end{figure*}

\subsection{Sod shock tube test}

The classical shock tube test is designed to demonstrate the code's shock-capturing capabilities.  We adopted 1D Cartesian coordinates, an ideal gas with $\gamma = 1.4$, and transmissive boundary conditions. The spatial domain is $x\in[0, 1]$, sampled with 100 grid points. The initial conditions were taken from \citet{Toro2009} and such that\begin{equation}
     (\rho, u, P)(x) =\left\{\begin{aligned}
        &(1.0, 0, 1.0) &\quad\textrm{if } x < 0.5\\
        & (0.125, 0, 0.1) &\quad\textrm{if } x > 0.5.
    \end{aligned}
    \right.
\end{equation}

This test is not particularly challenging for modern hydrodynamics codes but it is interesting because it exhibits all fundamental hydrodynamical waves. It is also well documented. The exact solution is known analytically and consists of a propagating rarefaction wave on the left side, a rightward-propagating contact discontinuity, and a rightward-propagating shock.
 
Figure~\ref{fig:tube} shows the analytic solution and the results from \heracles. The code correctly captures the different waves. The contact wave is the most diffuse, captured within approximately four grid points, whereas the shock wave spans around two grid points. The rarefaction wave is well described by our numerical scheme.

\subsection{2D Rayleigh-Taylor test}
\label{sect_appendix:2d_rt_test}

In Sect.~\ref{section:RT_3D} we presented results for the Rayleigh-Taylor instability test in 3D. Here, we consider a similar problem but in 2D and with different initial perturbations, which we took from \citet{Skinner2019}.

The fluid is contained in a box of size defined by $x\in[-0.5, 05]$, $z\in[-1, 1]$, and with $n_x=1024$ and $n_z=2048$. The gravitational acceleration is constant with $g_z =-0.5$ and the initial density of the stationary fluid is given by
\begin{equation}
    \rho(x, z)= \left\{\begin{aligned}
    & 2 \quad \textrm{if} \quad z \ge 0, \\
    & 1 \quad \textrm{otherwise.}
        \end{aligned}
\right.
\end{equation}
The initial pressure is given by the conditions for  hydrostatic equilibrium:
\begin{equation}
    P(z) = P_0 +\int_{-1}^z \rho g_z dz
,\end{equation}
with $P_0 = 10/7 + 1/4$. For this test we adopted an ideal gas with $\gamma$ = 1.4. At time $t=0$, the interface between the two fluids was perturbed according to
\begin{equation}
    h(x) = h_0 \cos(\kappa x)
,\end{equation}
where $h_0= 0.01$ and $\kappa = 4\pi$. The sharp interface is
then smoothed into an hyperbolic tangent profile with
characteristic length 0.005. The evolution of the fluid is tracked using a passive scalar $f_x$, which is set to zero where $z<0$ and 1 elsewhere.

\begin{figure}
    \centering
    \includegraphics[width=\columnwidth]{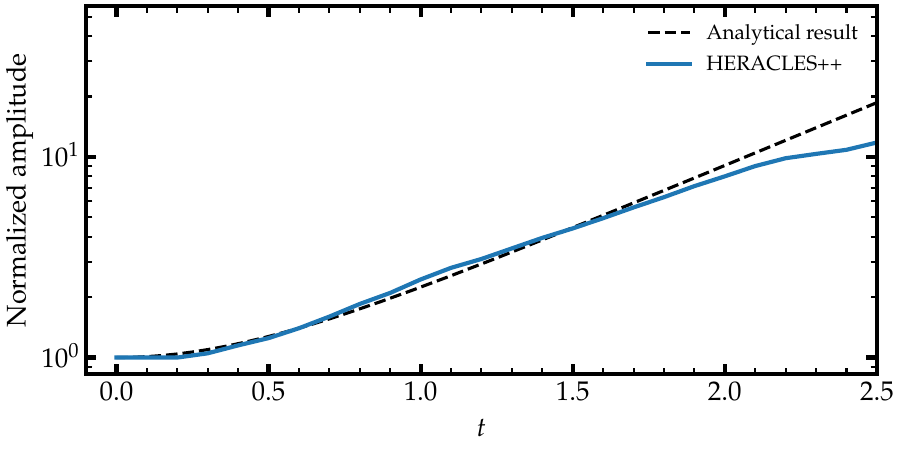}
    \caption{Amplitude of the interface perturbation for the single mode 2D Rayleigh-Taylor test and comparison with the analytical solution.}
    \label{fig:growth_rt2d}
\end{figure}

Figure~\ref{fig:RT_2D} shows the increasing mixing between the two fluids as time progresses, as well as the formation of the characteristic mushroom structures typical of the Rayleigh-Taylor instability. With the heavier fluid above, the slightest disturbance of the interface amplifies exponentially. Figure~\ref{fig:growth_rt2d} compares the evolution of the size of the mixing layer according to the analytical theory: $h(t) = h_0 \cosh{(\sqrt{Ag\kappa}\, t)}$, with $A = 1/3$, the Atwood number. The result shows a great agreement with the linear theory up to $t\sim$1.5, when the Rayleigh-Taylor structures start to develop.

\subsection{Liska-Wendroff implosion test}
\label{sect:liska}

\begin{figure}
    \centering
    \includegraphics[width=\columnwidth]{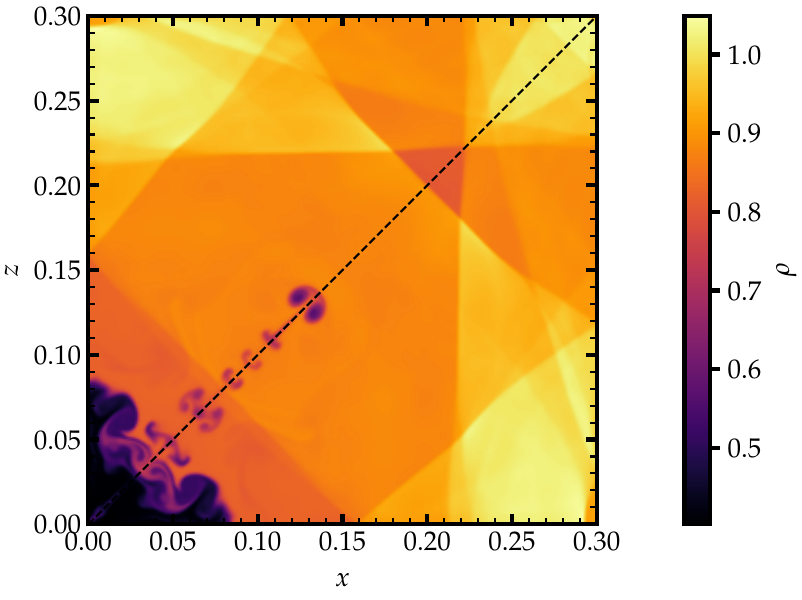}
    \caption{Results for the 2D implosion test. We show a color map of the density at $t=2.5$, with the $x=z$ symmetry axis in black dashed line. (see Appendix~\ref{sect:liska} for details.)}
    \label{fig:implosion}
\end{figure}

The 2D implosion test of \citet{Liska2003} consists of Sod-like initial conditions, but with a 45\,deg rotation. The boundary conditions are reflective in every direction and the size of the domain is $(x, z) \in[0, 0.3]$, which we cover with 1024 zones in each direction. We again adopted an ideal gas with $\gamma=1.4$. Following \citet{Liska2003}, the initial conditions of the stationary fluid are
\begin{equation}
    (\rho, P)(x, z)= \left\{\begin{aligned}
    & (1, 1) \quad &\textrm{if} \quad x+z \ge 0.015, \\
    &(0.14, 0.125)\quad &\textrm{otherwise.}
        \end{aligned}
\right. 
\end{equation}

This setup results in a shock moving away from the initial discontinuity, eventually reflecting off the boundaries and interacting with itself. The solution includes a low-density jet that travels along the symmetry axis $x=z$. This problem is very sensitive to the preservation of the reflection symmetry with the jet failing to form if symmetry is violated. 

We show the results for this test in Fig.~\ref{fig:implosion} and confirm the propagation of the jet along the symmetry axis. The distance traveled by the jet is a useful probe of the numerical diffusion. Indeed, at much higher resolutions, the jet is found to propagate further (not shown).

\subsection{Multidimensional Sedov blast wave}

In Sect.~\ref{section:sedov_3D} we present a 3D version of the off-center Sedov blast wave in the spherical coordinate. We also test the Sedov explosion in spherical 1D and 2D, with a centered explosion. We compare with the analytical solution.

For the 1D simulation, the domain is $r\in[0, 1]$ (which we sample with 400 grid points) and the explosion is initiated at $r=0$. The initial conditions are $\rho = 1, \quad u = 0$ and $\gamma=1.4$. $E=E_0=10^{-12}$ everywhere except at the point of explosion where $E=E_1=1$ at the point of the explosion. The boundary conditions are reflexive at the center and transmissive at the maximum radius. The 2D counterpart is started from the results of the 1D simulation at $t=0.3$, with the fluid properties copied across all polar directions. The spherical grid is such that $r\in[0.3, 1.3]$, $\theta \in [\pi/4, 3\pi/4]$, with 400 grid points in each direction.

\begin{figure}
    \centering
    \includegraphics[width=\columnwidth]{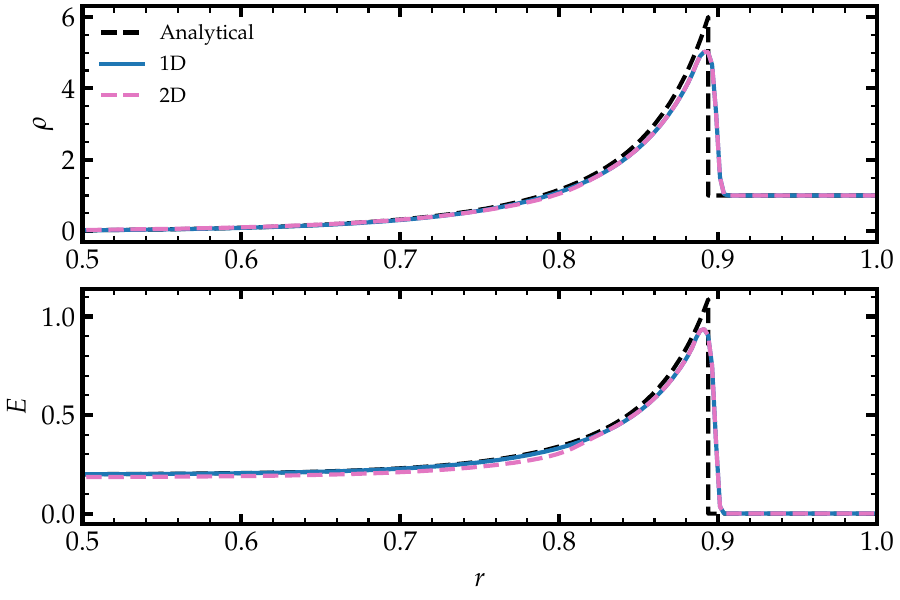}
    \caption{Comparison of the density (top) and total energy (bottom) at $t=0.7$ for the Sedov blast wave test performed with \heracles\ in 1D (blue) and 2D (pink; we show the result along $\theta=\pi/2$). We also show the analytic solution (black dashed line).}
    \label{fig:sedov_1d}
\end{figure}

\begin{figure}
    \centering
    \includegraphics[width=7cm]{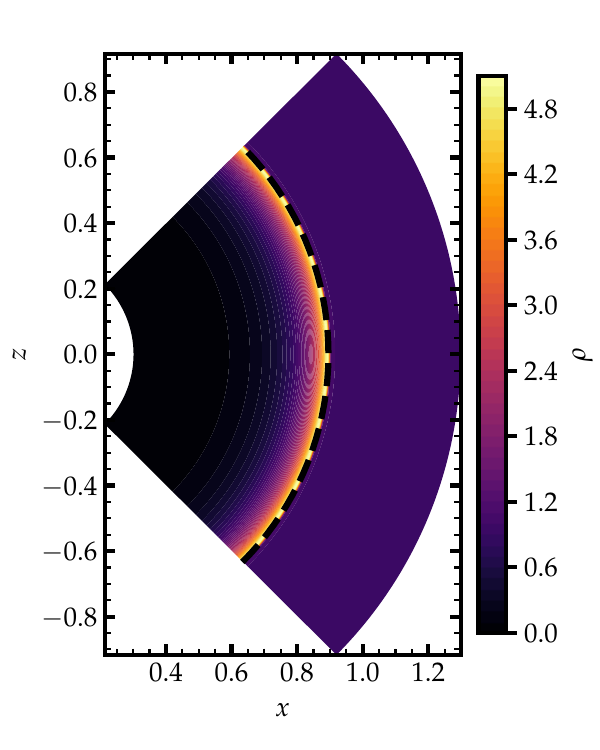}
    \caption{Color map of the density at $t=0.7$ for the 2D Sedov blast wave test (the analytic solution is shown as a black dashed line).}
    \label{fig:sedov_2d}
\end{figure}

Figure~\ref{fig:sedov_1d} compares the density and total energy at $t=0.7$ for the 1D and 2D simulations (we also added the analytical solution for each quantity). The position of the shock predicted by \heracles\ is similar in 1D and 2D and it also agrees with the analytical solution. Figure~\ref{fig:sedov_2d} shows the density at $t=0.7$. The shock remains spherical and its outer radius follows the analytical prediction (Eq.~\ref{eq:analytical_sedov}, with $\xi_0\approx 1.15$).

\end{appendix}

\end{document}